\documentclass[msc,oneside]{ubcthesis}
\usepackage{afterpage}
\usepackage{longtable}
\usepackage{graphicx}
\usepackage{pdflscape}
\usepackage[numbers,sort&compress]{natbib}
\usepackage{psfrag}
\usepackage[percent]{overpic}
\usepackage[unicode=true,
  linktocpage,
  linkbordercolor={0.5 0.5 1},
  citebordercolor={0.5 1 0.5},
  linkcolor=blue]{hyperref}
\usepackage{pdflscape}
\usepackage{graphicx}				
\usepackage{lipsum}
\usepackage{ntheorem}								
\usepackage{amssymb}
\usepackage{amssymb}
\usepackage{color}
\usepackage{amsmath}
\usepackage{amssymb}
\usepackage{float}
\usepackage{tikz}
\usepackage{tgtermes}
\usepackage[font={small,it}]{caption}
\usepackage{bm}
\usepackage{mathrsfs}
\usepackage{mathrsfs}
\usepackage{mathtools}
\newtheorem*{defn}{Definition}

\usepackage{mathtools}
\newcommand{\defeq}{\vcentcolon=}

\institution{The University Of British Columbia}
\faculty{The Faculty of Graduate and Postdoctoral Studies}
\institutionaddress{Vancouver}

\previousdegree{B.Sc., The University of New Mexico, 2014}

\title{Emergent Geometry Through Holomorphic Matrix Models}
\author{Stephen Pietromonaco}
\copyrightyear{}
\submitdate{\monthname\ \number\year} 
\program{Physics}

\renewcommand\thepart         {\Roman{part}}
\renewcommand\thechapter      {\arabic{chapter}}
\renewcommand\thesection      {\thechapter.\arabic{section}}
\renewcommand\thesubsection   {\thesection.\arabic{subsection}}
\renewcommand\thesubsubsection{\thesubsection.\arabic{subsubsection}}
\renewcommand\theparagraph    {\thesubsubsection.\arabic{paragraph}}

\floatstyle{ruled}
\newfloat{Program}{htbp}{lop}[chapter]

\begin{document}

\frontmatter


\maketitle                      
\begin{abstract}                
 
Over the years, deep insights into string theory and supersymmetric gauge theories have come from studying geometry emerging from matrix models.  In this thesis, I study the $\mathcal{N}=1^{*}$ and $\mathcal{N}=2^{*}$ theories from which an elliptic curve with modular parameter $\tau$ is known to emerge, alongside an elliptic function called the generalized resolvent into which the physics is encoded.  This is indicative of the common origin of the two theories in $\mathcal{N}=4$ SYM.  The $\mathcal{N}=1^{*}$ Dijkgraaf-Vafa matrix model is intrinsically holomorphic with parameter space corresponding to the upper-half plane $\mathbb{H}$.  The Dijkgraaf-Vafa matrix model 't Hooft coupling $S(\tau)$ has been previously shown to be holomorphic on $\mathbb{H}$ and quasi-modular with respect to $\text{SL}(2, \mathbb{Z})$.  The allowed $\mathcal{N}=2^{*}$ coupling is constrained to a Hermitian slice through the enlarged moduli space of the holomorphic $\mathcal{N}=1^{*}$ model.  

After explicitly constructing the map from the elliptic curve to the eigenvalue plane, I argue that the $\mathcal{N}=1^{*}$ coupling $S(\tau)$ encodes data reminiscent of $\mathcal{N}=2^{*}$.  A collection of extrema (saddle-points) of $S(\tau)$ behave curiously like the quantum critical points of $\mathcal{N}=2^{*}$ theory.  For the first critical point, the match is exact.  This collection of points lie on the \emph{line of degeneration} which behaves in a sense, like a boundary at infinity

I also show explicitly that the emergent elliptic curve along with the generalized resolvent allow one to recover exact eigenvalue densities.  At weak coupling, my method reproduces the inverse square root of $\mathcal{N}=2^{*}$ as well as the Wigner semi-circle in $\mathcal{N}=1^{*}$.  At strong coupling in $\mathcal{N}=1^{*}$, I provide encouraging evidence of the parabolic density arising in the neighborhood of the line of degeneration.  To my knowledge, the parabolic density has only been observed asymptotically.  It is interesting to see evidence that it may be exactly encoded in the other form of emergent geometry: the elliptic curve with the generalized resolvent.  
\end{abstract}

\tableofcontents                

\chapter{Acknowledgements}      

First and foremost, I am deeply indebted to my advisor, Dr. Joanna Karczmarek.  Without her tremendous technical assistance, intuition, support, and patience, this project could not have been completed.  I am also very grateful to Jamie Gordon, Malik Barrett, Javier Gonz\'{a}lez Anaya, Elliot Cheung, and Jake Bian for enlightening conversations and suggestions.  Finally, I am thankful to Konstantin Zarembo for gracious and helpful correspondence.

\mainmatter

\chapter{Introduction}

The notion of ``emergent geometry" in theoretical physics is the idea that the geometry underlying a theory is not provided as input, but rather ``emerges" from something more fundamental.  This has the manifest benefit of making the theory less arbitrary.  For example, one usually specifies a quantum field theory by choosing a spacetime manifold $M$, inventing an action depending on structures defined on $M$, and then attempting to extract information from Feynman's path integral prescription, the hope being that one can compute algebraic information like correlation functions.  Therefore, we have data in the form of correlation functions emerging from our choice of manifold $M$.  The obvious problem with this program is the arbitrary choice we must make for a spacetime geometry.  One can hope to turn this program around, and find geometrical structures emerging from more fundamental input data.

Two questions immediately arise.  First, what should we take as the initial input data?  This should be something concrete, and physically motivated.  Second, what sort of geometry will emerge, and what information will it provide?  Let us begin by tackling the first question.  One of the most basic, algebraic objects motivated by physics is a \emph{matrix model}.  This is the simplest possible quantum field theory: namely, a gauge theory defined on a 0-dimensional manifold.  A particular matrix model is determined by a choice of polynomial superpotential $W(x)$ and a certain collection $\Gamma$ of $N \times N$ matrices $\Phi$ over which to integrate.  The partition function is then given by

\begin{equation}
\mathcal{Z}_{N}(\Gamma, g_{s}) \defeq \frac{1}{\mathcal{C}} \int_{\Gamma} d\Phi e^{-\frac{1}{g_{s}} \, \textnormal{Tr} W(\Phi)},
\end{equation}
where $\mathcal{C}$ is a normalization constant and $g_{s}$ is the coupling, playing an analogous role to $\hbar$ in quantum mechanics.  Notice that in a general gauge theory, the action requires integrating the Lagrangian density over the manifold, but here, with a zero-dimensional spacetime, this integral is replaced by the trace.  This is a concrete and computable quantity, certainly in comparison to richer quantum field theories.  However, such a model appears to be completely devoid of any elegant geometry at first sight.  One profoundly important insight has been that by taking the 't Hooft limit of certain matrix models, one can extract quite rich geometrical structures.

To tackle the second question, we ask what kinds of geometries do these models encode?  The first example of geometry emerging from matrix models is through \emph{eigenvalue densities}.  This originated with Eugene Wigner, who considered Hermitian matrix models with Gaussian actions \cite{wigner_characteristic_1993}.  Hermitian matrix models are matrix models where the eigenvalues distribute themselves on the real axis.  In such models with Gaussian actions, one can study the statistical distribution of the eigenvalues in the 't Hooft limit.  Wigner noticed that in this limit, the eigenvalues are distributed continuously over a path, called a cut.  The density of this distribution is a semi-circle.  This is Wigner's famous semi-circular distribution and is the first instance of geometry emerging from matrix models.  It is rather remarkable that such a rigid shape can emerge from a path integral over matrices.  In this thesis, I will study models which coincide with Wigner's Gaussian model in certain limits, and I exactly recover the semi-circular distributions.  In a related fashion, I will also exactly recover the inverse square-root of Shenker and Douglas \cite{douglas_dynamics_1995}.  Recently, people have been interested in matrix models in which a parabolic eigenvalue distribution emerges.  This density indicates a multi-matrix model in a commuting phase, and to my knowledge, has only been observed numerically, or by employing approximations.  I will provide partial evidence in support of an analytical method to recover the parabolic density.

The second form of emergent geometry is the \emph{algebraic varieties} which arise from the equations of motion of certain matrix models.  In \cite{dijkgraaf_matrix_2002} it was realized that one-matrix models with a polynomial superpotential determined hyperelliptic curves.  This was studied in detail in \cite{lazaroiu_holomorphic_2003-1}, which motivated our following chapter.  Also in \cite{dijkgraaf_matrix_2002} it was shown that these matrix models play a crucial role in the geometric transitions in topological string theories on Calabi-Yau manifolds.  In this thesis I will be concerned with the well-known solutions to $\mathcal{N}=1^{*}$ and $\mathcal{N}=2^{*}$ theories in which an elliptic curve emerges.  It is quite astounding that something like an algebraic variety can emerge from something seemingly so structureless like a matrix model.

There is also a third form of emergent geometry from matrix models: the classical equations of motion of a multi-matrix model may define a non-commutative (fuzzy) geometry.  This is usually referred to as a \emph{non-commutative background geometry}.  For example, one might have a three-matrix model involving a path integral over three matrices $\Phi_{1}$, $\Phi_{2}$, and $\Phi_{3}$ in which the classical equations of motion constrain the matrices to satisfy

\begin{equation}
[\Phi_{1}, \Phi_{2}] = i \Phi_{3}, 
\end{equation}
in addition to the two cyclic permutations \cite{oconnor_critical_2013}.  In this case, the background geometry is a non-commutative two-sphere.  

I have reviewed three manifestations of geometry encoded into matrix models: the distributions of eigenvalues, the semi-classical algebraic curves, and the non-commutative background geometries.  A connection was recently established between two of these \cite{berenstein_multi-matrix_2009,ydri_remarks_2015,oconnor_critical_2013,delgadillo-blando_geometry_2008} studying multi-matrix models which enjoy a commuting phase given by a parabolic eigenvalue density, as well as a non-commuting phase.  In the low-temperature limit, as one passes to the non-commuting phase, the non-commutative sphere emerges.  The conclusion is that a geometry emerges as a system ``cools"; perhaps the geometry in our universe emerged post-Big Bang in an analogous (albeit, much more complex) way.  So these particular models link emergent eigenvalue densities and emergent non-commutative geometries.  

\emph{One aim of my thesis is to describe how this story relates to an emergent algebraic curve.  I will attempt to recover eigenvalue densities from this curve, which will connect all three forms of emergent geometry discussed above.}  As mentioned, I will study models which are intimately related to an elliptic curve.  The physics will determine a modular parameter $\tau$, as well as a particular elliptic function on the curve.  From these two bits of physically determined data, I will provide an exact algorithm for recovering eigenvalue densities, which to my knowledge, is not explicit in the literature.  This method will recover not only Wigner's semi-circle and Shenker's inverse square-root, but also a hint of the parabolic eigenvalue density.  So it appears that in certain cases, the eigenvalue densities are actually encoded into an emergent algebraic variety.

The discussion will rely on a generalization of an Hermitian matrix model, allowing for much richer structure.  These are known as \emph{holomorphic matrix models} and I introduce them in the next chapter.  In broad brush, these models allow for eigenvalues to be supported on more general paths in the complex plane, whereas Hermitian models have eigenvalues constrained to the real axis.  The emergent algebraic curves are varieties over $\mathbb{C}$, not $\mathbb{R}$, so this is very natural.  Holomorphic matrix models explore in full the relevant portion of the moduli spaces of these curves.  Their Hermitian siblings, merely explore a ``slice" through the relevant moduli space.  Having access to a larger region within a moduli space means that one can potentially avoid critical points.  In a Hermitian matrix model, there exists the possibility of complete termination upon colliding with a critical point.  However, if it's possible to embed the theory into a holomorphic model, one can avoid the critical point and explore the bulk of the moduli space.  I will attempt a maneuver of this sort in this thesis.  In particular, the $\mathcal{N}=2^{*}$ theory is well-known to have an infinite sequence of critical points accumulating at strong coupling.  It is also well-known that at weak coupling an emergent elliptic curve offers an exact solution of the theory, but beyond the first critical point the Hermitian slice degenerates completely.  I hope to argue that by employing holomorphic matrix models, we may by-pass this first critical point, continuing to a region of the moduli space which may shed light on strong coupling behavior.  More generally, the fundamental question I will consider is the following.

\begin{center}
\textit{Given a Hermitian matrix model in the 't Hooft limit whose weak coupling region embeds naturally along the Hermitian slice up to the first critical point, can the geometrical solution using an elliptic curve detect the strong coupling region?}
\end{center}

\noindent In Chapters 5 and 6, I answer partially in the affirmative for the $\mathcal{N}=2^{*}$ theory as well as the Hermitian model studied in \cite{berenstein_multi-matrix_2009,ydri_remarks_2015,oconnor_critical_2013,delgadillo-blando_geometry_2008}, respectively.

Earlier I indicated that the program was to \emph{choose} a matrix model as input, from which richer structure may emerge.  This is a good general philosophy, however the matrix models considered in my thesis actually arise from attempts to answer difficult questions in quantum field theory.  In quantum field theories, computing physical observables like the partition function, or correlation functions, requires computing Feynman's path integral.  In general, the path integral is understood only as a schematic or heuristic device.  In certain special cases however, the infinite dimensional path integral ``localizes" to something manageable.  For example, it might localize to an ordinary integral over a concrete space, or a matrix model, or perhaps even a sum.  Compared to an infinite dimensional path integral, a matrix model is a relatively computable quantity.  It is in this sense that an impenetrable observable in a quantum field theory may be computed by considering a matrix model.  More specifically, there will be a certain class of observables called a subsector which can be exactly computed.  Using matrix models to compute these observables will not be the main focus of my thesis.  Rather, I will be studying the matrix models themselves.  Nevertheless, it is important to understand where the objects one is studying came from.  I provide a very brief outline of this.

One of the most fundamental quantum field theories is $\mathcal{N}=4$ super Yang-Mills (SYM) in four dimensions.  This theory is superconformally invariant, even quantum mechanically, and it enjoys the mysterious $S$-duality, which seems to identify each physically distinct parameter value with an elliptic curve.  This theory has two unique massive deformations preserving $\mathcal{N}=1$ and $\mathcal{N}=2$ supersymmetry: these resulting theories are known as $\mathcal{N}=1^{*}$ and $\mathcal{N}=2^{*}$, respectively.  

\subsection*{The $\mathcal{N}=1^{*}$ Theory}

In general, \, $\mathcal{N}=1$ supersymmetric gauge theories on $M=\mathbb{R}^{4}$ contain a ``holomorphic subsector."  This is a special class of physical observables which are constrained to depend holomorphicaly on the parameters of the theory.  More specifically what this means is that given such a theory, one can compute an \emph{effective superpotential} $W_{\textnormal{eff}}$, which is a holomorphic function of the parameters.  In turn, the vacuum expectation values of the special observables are given by differentiating $W_{\textnormal{eff}}$ with respect to these parameters.  All observables that can be obtained in this manner define the holomorphic subsector.  The question immediately becomes, how does one compute $W_{\textnormal{eff}}$?  Dijkgraaf and Vafa \cite{dijkgraaf_perturbative_2002} have proposed a solution that works in particular for the $\mathcal{N}=1^{*}$ theory.  

The $\mathcal{N}=1^{*}$ theory enjoys an extremely rich classical vacuum structure.  In fact, each classical vacuum will be given by a non-commutative background geometry of the form described earlier.  The vacua are discrete, and depend on the amount of symmetry broken in the original gauge group.  There will be a different effective superpotential for each different vacuum.  The Dijkgraaf-Vafa prescription is that one should consider a holomorphic matrix model expanded around the relevant vacuum.  They conjecture that the 't Hooft limit of this matrix model computes observables in the field theory for \emph{finite} $N$ in the holomorphic subsector.  In particular, one can exactly compute the effective superpotential $W_{\textnormal{eff}}$ in this vacuum which will depend on a modular parameter $\tau$.  The holomorphic observables will arise by differentiating $W_{\textnormal{eff}}$ with respect to $\tau$.  

This conjecture is in the same spirit described above: an impenetrable computation in quantum field theory localizes to something manageable and computable.  In this case, the localization refers to considering a matrix model as a fluctuation about a particular classical vacua.

\subsection*{The $\mathcal{N}=2^{*}$ Theory}

One of the landmark achievements recently in theoretical physics, is the realization by Pestun \cite{pestun_localization_2012} that certain observables in $\mathcal{N}=2$ supersymmetric gauge theories are exactly computable on a spacetime $M=S^{4}$.  This remarkable method is known as \emph{localization}.  Roughly speaking, the field content of $\mathcal{N}=2$ SYM contains a scalar field $\Phi$ transforming in the adjoint representation of a gauge group, whose vacuum expectation value (VEV) is

\begin{equation}
\langle \Phi \rangle = \textnormal{diag}(a_{1}, \ldots, a_{N}).
\end{equation}
The eigenvalues $\{a_{I}\}$ of the adjoint scaler field VEV are called Coulomb moduli as they parameterize the Coulomb branch moduli space.  Pestun's localization reduces the computation of certain observables to finite-dimensional matrix integrals over the Coulomb moduli.  For instance, the partition function of $\mathcal{N}=2$ super Yang-Mills compactified on $S^{4}$, localizes to a finite-dimensional matrix model.  The partition function of $\mathcal{N}=2^{*}$ on $S^{4}$ also localizes to a matrix model.  I will study this matrix model by letting the radius of $S^{4}$ go to infinity (the decompactification limit) and simultaneously taking the 't Hooft limit.  It is precisely in such a scenario where $S^{4}$ can be approximately identified with $\mathbb{R}^{4}$ and we may see connections to the emergent geometry of $\mathcal{N}=1^{*}$.  
\vskip2ex

I make one final informal remark about the mathematical analog to localization in physics.  Ideally, one would like to be able to integrate any differential form over any smooth manifold $M$.  This is a non-trivial problem given that in general, actually evaluating an integral is merely inspired guesswork.  If however, one has a group action on the manifold, there is a special class of forms called equivariant forms whose integral over $M$ reduces to a sum over indices at the fixed points of the group action.  This is known as Atiyah-Bott localization in mathematics.  But notice how analogous this is to our physical problem in quantum field theory: observables require computing a path integral, even whose \emph{definition}, much less exact value eludes us.  However, in certain theories, we have a special class of observables (a subsector) which may be computed exactly in an easy way using some ``localization" method.  In $\mathcal{N}=2^{*}$ this is \emph{the} localization of Pestun, while in $\mathcal{N}=1^{*}$, this is the Dijkgraaf-Vafa conjecture using holomorphic matrix models.  

\emph{Again, the quantum field theory questions will not be the main focus of my thesis, but rather the matrix models themselves.}  However, I hope that it is enlightening to understand why I am studying these matrix models, and where they came from.  Moreover, certain results can be interpreted as a ``shadow" of some structure in the original theory before deformation or localization.  For instance, the elliptic curves emerging from the mass-deformed matrix models seem to be an artifact of their origin in $\mathcal{N}=4$ SYM.  Finally, and most importantly, perhaps some matrix model result might help in understanding quantum field theory questions in future work.

\chapter{Introduction to Holomorphic Matrix Models}

This chapter is modeled on Calin Lazaroiu's excellent paper \cite{lazaroiu_holomorphic_2003-1} and contains \emph{no} original results.  The goal is to introduce the foundations of holomorphic matrix models which will pervade the later chapters.  

\section{Definition of the Model at Finite $N$}

I begin by considering $\textnormal{Mat}_{N}(\mathbb{C})$, the space of $N \times N$ matrices with complex entries.  This is an $N^{2}$-dimensional manifold identified with $\mathbb{C}^{N^{2}}$.  Let the subset $\mathcal{D} \subset \textnormal{Mat}_{N}(\mathbb{C})$ be defined to contain all matrices which are diagonalizable.  In other words, for all $\Phi \in \mathcal{D}$, there exists a general linear transformation $S \in \textnormal{GL}_{N}(\mathbb{C})$ such that $\Phi$ becomes diagonal upon conjugation by $S$

\begin{equation}
S \Phi S^{-1} = \textnormal{diag}(\lambda_{1}, \ldots, \lambda_{N}),
\end{equation}
where the $\lambda_{I}$ are the eigenvalues of $\Phi$.  The subset $\mathcal{D}$ constitutes an open submanifold of $\textnormal{Mat}_{N}(\mathbb{C})= \mathbb{C}^{N^{2}}$.  One can define the spectrum $\sigma(\Phi)$ of a matrix $\Phi \in \mathcal{D}$ to be simply the set of eigenvalues

\begin{equation}
\sigma(\Phi) \defeq \{ \lambda_{1}, \ldots, \lambda_{N}\}.
\end{equation}
Let $\Gamma$ denote a connected, non-compact, boundary-less submanifold of $\mathcal{D}$, which is \emph{half-dimensional}, i.e.,

\begin{equation}
\textnormal{dim}_{\mathbb{R}} \Gamma = \textnormal{dim}_{\mathbb{C}} \mathcal{D} = N^{2}.
\end{equation}
In other words, the real dimension of $\Gamma$ is precisely the complex dimension of $\mathcal{D}$; it is in this sense that $\Gamma$ is a half-dimensional ``real slice" through $\mathcal{D}$.  

A polynomial superpotential is defined by the following degree-$(n+1)$ polynomial ($n \geq 2$) with complex coefficients

\begin{equation}
W(x) = \sum_{I=0}^{n+1} u_{I}x^{I}.
\end{equation}
We will soon see that for my purposes, only the critical points of $W(x)$ will be relevant and thus, all polynomial superpotentials which differ by a constant will be identified.  With these basic notions in hand, I can now give a definition of a holomorphic matrix model.
\vskip1ex

\begin{defn}
Given a choice of superpotential $W(x)$, a choice of $\Gamma \subset \mathcal{D}$ as described above, and a choice of $N$, a \textbf{holomorphic matrix model} is defined by the partition function

\begin{equation} \label{eqn:partfuncLaz}
\mathcal{Z}_{N}(\Gamma, g_{s}) \defeq \frac{1}{\mathcal{C}} \int_{\Gamma} d\Phi e^{-\frac{1}{g_{s}} \, \textnormal{Tr} W(\Phi)},
\end{equation}
where $\mathcal{C}$ is a normalization constant, $d\Phi = \wedge_{I,J} \Phi_{I,J}$ is the familiar matrix integration measure, and $g_{s}$ is a coupling constant.   
\end{defn}
A holomorphic matrix model is a path integral over a certain half-dimensional ``contour" within the space of complex matrices.  However, it is much nicer to work with the eigenvalue representation of a matrix model.  Therefore we want to translate the above definition into an integral over the eigenvalues of the matrix.  Let $\mathbb{C}$ be called the \emph{eigenvalue plane} and let $\gamma : \mathbb{R} \to \mathbb{C}$ be an open, immersed path in this plane, without self-intersections.  We can use this path $\gamma$ to define the subspace of $\mathcal{D}$

\begin{equation}
\Gamma(\gamma) \defeq \{ \Phi \in \mathcal{D} \big| \sigma(\Phi) \subset \gamma\},
\end{equation}
where I slightly abuse notation and identify the path $\gamma$ with its image in $\mathbb{C}$.  Notice that since $\gamma \subset \mathbb{C}$ is half-dimensional (i.e., the real dimension of $\gamma$ coincides with the complex dimension of $\mathbb{C}$), it follows that $\Gamma(\gamma)$ is in turn also half-dimensional inside of $\mathcal{D}$.  Lazaroiu \cite{lazaroiu_holomorphic_2003-1} proves that the original partition function (\ref{eqn:partfuncLaz}) is actually gauge invariant, with the gauge orbits being given by the complex homogeneous space $H \defeq \textnormal{GL}_{N}(\mathbb{C})\big/ (\mathbb{C}^{*})^{N}$.  Denote by $\textnormal{hvol}(H)$ the volume of the space $H$.  Using this, we can choose the normalization $\mathcal{C}$ to cancel all unwanted factors

\begin{equation}
\mathcal{C} = (-1)^{N^{2}(N-1)/2}\frac{1}{N!} \textnormal{hvol}(H).
\end{equation}
With this normalization, we can give a very nice expression for the eigenvalue representation of a holomorphic matrix model

\begin{equation} \label{eqn:eigvalrep}
\mathcal{Z}_{N}(\gamma, g_{s}) = \int_{\gamma} d \lambda_{1} \cdots \int_{\gamma} d \lambda_{N} \prod_{I \neq J} (\lambda_{I} - \lambda_{J})^{2}e^{-\frac{1}{g_{s}} \sum_{J=1}^{N} W(\lambda_{J})},
\end{equation}
where we simply define $\mathcal{Z}_{N}(\gamma, g_{s}) \defeq \mathcal{Z}_{N}(\Gamma(\gamma), g_{s})$.  When changing variables to the eigenvalue representation, it is the Jacobian which gives rise to the Vandermonde determinant $\Delta(\lambda) \defeq \prod(\lambda_{I}-\lambda_{J})$.  

At this stage, it is obvious what a holomorphic matrix model is: we must provide the data of a polynomial superpotential $W(x)$ as well as a suitable path $\gamma$, and the model will be defined as a path integral over all matrices whose eigenvalues lie entirely within $\gamma$.  If we choose $\gamma$ to be the real axis, we recover the familiar Hermitian matrix models.  Similarly, choosing $\gamma$ to be the imaginary axis or the unit circle, gives rise to an anti-Hermitian matrix model or unitary matrix model, respectively.  In this sense, holomorphic matrix models generalize these models by allowing the eigenvalues to be supported on more general paths in the complex plane.  

Section 2 of Lazaroiu's paper deals with convergence issues.  I will not provide the details here, as they will not constrain us greatly.  Briefly, depending on the polynomial $W(x)$, the path $\gamma$ needs to asymptote at infinity to certain specific sectors of the complex plane.  For example, if we choose $W(x)$ to have odd degree, then choosing $\gamma$ to be the real axis will not work.  This is an illustration of the familiar fact that Hermitian matrix models are not well-defined for polynomial superpotentials of odd degree, since the real part of the potential cannot be bounded from below along the real axis.  In Section 2 of his paper \cite{lazaroiu_holomorphic_2003-1}, Lazaroiu provides the constraints on $\gamma$ from $W(x)$.

\section{The 't Hooft Limit and the Emergence of a Hyperelliptic Curve}

Let $\gamma$ be a path in $\mathbb{C}$ as described in the previous section.  Following Lazariou, let $s$ denote a length coordinate on the path.  Thus, $\lambda(s)$ shall be the parameterization of the path, and for notational ease, let $\lambda_{I} = \lambda(s_{I})$.  Since the matrix model eigenvalues are constrained to live in $\gamma$, we want to introduce an eigenvalue density along the path.  For finite $N$, this is simply a sum of delta functions

\begin{equation}
\rho(s) \defeq \frac{1}{N} \sum_{I=1}^{N} \delta(s - s_{I}),
\end{equation}
with normalization condition

 \begin{equation} \label{eqn:normalization}
 \int_{-\infty}^{\infty} ds \rho(s) =1.
 \end{equation}
Note that the range of the integral is from $-\infty$ to $\infty$.  Following the usual matrix model constructions, define the resolvent

\begin{equation}
\omega(x) \defeq \frac{1}{N} \sum_{I=0}^{N} \frac{1}{x - \lambda_{I}}.
\end{equation}
In the eigenvalue representation of the partition function (\ref{eqn:eigvalrep}), we integrated over the eigenvalues $\lambda_{I}$.  Now, make a change of variables $\lambda_{I} = \lambda(s_{I})$, and rewrite the partition function in terms of the $s_{I}$.

\begin{equation}
\mathcal{Z}_{N}(\gamma, g_{s}) = \int_{\gamma} ds_{1} \cdots \int_{\gamma} ds_{N} \prod_{I=1}^{N} \dot{\lambda}(s_{I}) \prod_{I \neq J}\big(\lambda(s_{I})-\lambda(s_{J})\big)^{2}e^{-\frac{1}{g_{s}}\sum _{I=1}^{N} W(\lambda(s_{I}))}.
\end{equation}
It is illustrative to introduce an effective action $S_{\textnormal{eff}}$,

\begin{equation} \label{eqn:effact}
S_{\textnormal{eff}}(s_{1}, \ldots, s_{N}) \defeq \sum_{J=1}^{N} W(\lambda_{J})- 2 g_{s} \sum_{I \neq J} \log(\lambda_{I} - \lambda_{J}) - g_{s} \sum_{I=1}^{N} \log \dot{\lambda}_{I},
\end{equation}
such that the total partition function can now be written as

\begin{equation}
\mathcal{Z}_{N}(\gamma, g_{s}) = \int_{\gamma} ds_{1} \cdots \int_{\gamma} ds_{N} \, e^{-\frac{1}{g_{s}} \, S_{\textnormal{eff}}}.
\end{equation}
One can then vary the effective action with respect to parameter $\lambda_{I} = \lambda(s_{I})$ which gives the equation of motion

\begin{equation} \label{eqn:eqnmot1}
 W'(\lambda_{I}) -2g_{s} \sum_{J \neq I} \frac{1}{\lambda_{I}-\lambda_{J}} - g_{s} \frac{\ddot{\lambda}_{I}}{\dot{\lambda}_{I}^{2}}=0.
\end{equation}
We see here that the equation of motion depends on $W'$, not $W$, further emphasizing that superpotentials differing by a constant should be physically identified.  Given the definition of the resolvent, one can show that the equation of motion above is equivalent to

\begin{equation} \label{eqn:eqmot2}
\omega(x)^{2} +\frac{1}{g_{s} N} W'(x)\omega(x) + \frac{1}{4 (g_{s} N)^{2}}f(x) + \frac{1}{N} \omega'(x) + \frac{1}{N^{2}} \sum_{I=1}^{N} \frac{\ddot{\lambda}_{I}}{\dot{\lambda}_{I}^{2}}=0,
\end{equation}
where the polynomial $f(x)$ is defined as

\begin{equation} \label{eqn:fsum}
f(x) \defeq 4g_{s} \sum_{I=1}^{N} \frac{W'(x) - W'(\lambda_{I})}{x-\lambda_{I}}.
\end{equation}
Notice that the equation of motion (\ref{eqn:eqmot2}), as written, is a differential equation for the resolvent $\omega(x)$.  If we let $N \to \infty$, it seems like we can discard the term with the derivative, leaving simply an algebraic equation.  The downside to this appears to be that we will lose almost every term above, leaving the trivial condition $\omega(x)^{2} = 0$.  However, we can take the large-$N$ limit in specifically such a way as to keep most of the relevant terms above.  Let $N \to \infty$, and simultaneously let $g_{s} \to 0$, while keeping fixed the 't Hooft coupling which is defined as

\begin{equation}
S = g_{s} N.
\end{equation}
This is known as the \emph{'t Hooft limit} and it is one of the most important limits in a gauge theory.  It is precisely the limit in which many gauge theories are conjectured to find intimate ties to string theory.

Taking the 't Hooft limit of (\ref{eqn:eqmot2}), the last two terms drop out entirely, leaving the purely algebraic equation

\begin{equation} \label{eqn:thooftmot}
\omega(x)^{2} +\frac{1}{S} W'(x)\omega(x) + \frac{1}{4 S^{2}}f(x)=0.
\end{equation}
There is one subtlety here (see Lazariou for details).  In taking the 't Hooft limit, we replace averaged quantities with their expansions in $\tfrac{1}{N}$.  For example, we have $\langle \rho(s) \rangle = \rho_{0}(s) + \mathcal{O}(\tfrac{1}{N})$.  Thus, when defining the resolvent and $f(x)$ in the 't Hooft limit, it is done with respect to the eigenvalue density $\rho_{0}(s)$, as follows,

\begin{equation} \label{eqn:resolvv}
\omega(x) = \int_{-\infty}^{\infty} ds \frac{\rho_{0}(s)}{x - \lambda(s)},
\end{equation}
\begin{equation} \label{eqn:quantdef}
f(x) = 4 S \int_{-\infty}^{\infty} ds \rho_{0}(s) \frac{W'(x) - W'(\lambda(s))}{x - \lambda(s)}.
\end{equation}
There's a constraint on $f(x)$ coming from the normalization of the eigenvalue density.  One can see that the leading coefficient must be $4S(n+1)u_{n+1}$, where we recall that $u_{n+1}$ is the leading coefficient of $W(x)$, \cite{lazaroiu_holomorphic_2003-1}.

A strong word of warning is due at this point.  The eigenvalue density $\rho_{0}(s)$ is actually \emph{complex-valued}.  This will become important later in the thesis.  The details can all be found in Lazaroiu \cite{lazaroiu_holomorphic_2003-1}, noting that he renames the large-$N$ quantities $\omega_{0}(x)$ and $f_{0}(x)$.

Defining the function
\begin{equation}
y(x) = 2 S \omega(x) + W'(x), 
\end{equation}
one can show that the equation of motion in the 't Hooft limit (\ref{eqn:thooftmot}) is equivalent to

\begin{equation} \label{eqn:hypellcurv}
\boxed{y^{2} - W'(x)^{2} + f(x) =0.} 
\end{equation}

\begin{defn}
A smooth \textbf{hyperelliptic curve of genus $\bm{g}$} is an algebraic curve over $\mathbb{C}$ given by $y^{2} - F(x)=0$ where $F$ is a polynomial of even degree $2n=2(g+1) \geq 4$ with $2n$ distinct roots and complex coefficients.  For our purposes, a \textbf{maximally singluar} hyperelliptic curve is the singular curve of the form $y^{2} - H(x)^{2}=0$ for a polynomial $H$ of degree $n \geq 2$ with distinct roots.
\end{defn}

\noindent Since $W(x)$ is a degree-$(n+1)$ polynomial, $W'(x)^{2}$ is a degree-$2n$ polynomial.  The polynomial $f(x)$ is of degree-$n$, so we see that the algebraic equation (\ref{eqn:hypellcurv}) is exactly that of a hyperelliptic curve!  Thus, in the 't Hooft limit, the equation of motion of the matrix model gives rise to a hyperelliptic curve, which is thought to double-cover the eigenvalue plane compactified to $\mathbb{P}^{1}$, branched over the $2n$ zeros of $W'(x)^{2}-f(x)$.  This is an example of an emergent algebraic curve, as mentioned in the Introduction.  In algebraic geometry, hyperelliptic curves have a number of moduli determining the complex structure.  If we choose to consider a Hermitian matrix model, we would be constraining ourselves to a ``slice" through the moduli space of hyperelliptic curves.  It is the holomorphic models which explore the full moduli space of emergent structures.

\section{Physical Discussion of Classical and 't Hooft Limits}

The previous section developed the structure of the theory, but I have yet to provide any physical interpretation or intuition.  In this section, I describe the physics of a holomorphic matrix model in the 't Hooft limit \cite{dijkgraaf_matrix_2002,dijkgraaf_geometry_2002}, which was introduced above.  But first, I discuss the \emph{classical limit} $g_{s} =0$.  This is analogous to taking the $\hbar =0$ limit in quantum mechanics.  Mathematically, this limit corresponds to a maximally singular hyperelliptic curve.

\section*{The Classical Limit}

As mentioned, the classical limit of the theory corresponds to fixing $g_{s} =0$.  In taking the 't Hooft limit, we let $g_{s} \to 0$, but this is fundamentally different.  If one has already taken the 't Hooft limit, the classical limit can be recovered by requiring $S=0$.  In this limit, we see from (\ref{eqn:effact}) that the effective action is given by

\begin{equation}
S_{\textnormal{eff}} = \sum_{J=1}^{N} W(\lambda_{J}),
\end{equation}
which is simply the action for $N$ non-interacting eigenvalues, under the influence of a polynomial potential.  In addition, from (\ref{eqn:eqnmot1}) the equation of motion reduces to

\begin{equation}
W'(\lambda_{I}) = 0,
\end{equation}
which is the statement that the eigenvalues live precisely at the extrema of the polynomial potential.\footnote{For $n \geq 1$ extrema, one must specify \emph{filling fractions} which are $n$ rational numbers satisfying $q_{1} + \cdots + q_{n}=1$ which specify the fraction of eigenvalues in each extrema.}  Most importantly, we see from (\ref{eqn:fsum}) that the polynomial $f(x)$ vanishes when $g_{s}=0$.  That is to say, the holomorphic matrix model in its classical limit is described by a maximally singular hyperellpitic curve \cite{dijkgraaf_matrix_2002},

\begin{equation}
y^{2} = W'(x)^{2}.
\end{equation}
The singularities of the hyper elliptic curve coincide with the extrema of $W(x)$.  The physical interpretation is that we've distributed the $N$ eigenvalues into the extrema of the polynomial, and since there are no interaction terms, they simply reside at the extrema.  As noted in \cite{dijkgraaf_matrix_2002}, working in the holomorphic setting, one does not distinguish between minima and maxima; all extrema are saddle points.

\section*{The 't Hooft Limit}

Turning on the coupling $g_{s}$, the effective action retains its form in (\ref{eqn:effact}).  The first logarithmic term in the effective action acts to repel the eigenvalues from one another.  This provides the impetus for the eigenvalues to depart from their positions at the extrema of the polynomial.  Thus, the two physical phenomena present are an attraction of each eigenvalue to the extrema of $W(x)$, as well as a mutual repulsion of the eigenvalues.  As we have seen, in the 't Hooft limit the only free physical parameter is the coupling $S=g_{s} N$.  It is clear from the equations that $S=0$ recovers the classical limit, and singular algebraic curve.  If $S \neq 0$, the eigenvalues will spread out into a continuous cut.  Recalling that the eigenvalues are constrained to live in the path $\gamma$ these cuts are completely supported within $\gamma$.  These cuts are precisely the branch cuts of the hyperelliptic curve (\ref{eqn:hypellcurv}) and the endpoints correspond to the branch points.  Let $C_{1}, \ldots, C_{n}$ denote the $n$ cuts.  Important quantities are the periods of the Riemann surface given as certain contour integrals around the cuts,

\begin{equation}
S_{i} = \frac{1}{2 \pi i} \oint_{C_{i}} y(x) dx,
\end{equation}
where $S=S_{1} + \cdots + S_{n}$.  Note that these periods are invariants of a Riemann surface with a given complex structure.  They are holomorphic functions of the complex moduli of the surface.\footnote{For the remainder of my thesis, the matrix models will have only one cut, in which case $S_{1} =S$.}

\section{Summary and Outlook}

The most fundamental properties of holomorphic matrix models are that they have access to a larger moduli space than their Hermitian counterparts, and the physical couplings (like $S_{i}$) emerge as holomorphic functions of these complex moduli.  In the remainder of this thesis I study the $\mathcal{N}=1^{*}$ matrix model, which is intrinsically holomorphic.  Though it is of a slightly more complicated form than the models introduced here, it shares these same fundamental properties and in the coming chapters, I attempt to find interesting phenomena encoded into the $\mathcal{N}=1^{*}$ moduli space as well as possible connections to $\mathcal{N}=2^{*}$.

\chapter{Introduction to the $\mathcal{N}=1^{*}$ and $\mathcal{N}=2^{*}$ Theories}

The $\mathcal{N}=1^{*}$ and $\mathcal{N}=2^{*}$ theories are \emph{quantum field theories} defined as massive deformations of $\mathcal{N}=4$ super Yang-Mills (SYM) in four dimensions.  I begin this chapter by briefly describing this theory and defining the massive deformations.  The remaining two sections will be devoted to studying the $\mathcal{N}=1^{*}$ and $\mathcal{N}=2^{*}$ theories, respectively, in detail.  The exact solutions of both models requires considering certain matrix models.  In both cases, the equations of motion of the matrix models determine a particular elliptic curve with modular parameter $\tau$, as well as a certain elliptic function on the curve.  The physical parameters of the matrix model are encoded into $\tau$ as essentially a change of variables.  The similarity of the two solutions indicates that perhaps through the philosophy of \emph{holomorphic} models, an enlarged moduli space may uncover connections between the theories.  Such connections will occupy later chapters.  We will see in the next section that $\mathcal{N}=4$ SYM is intrinsically connected to an elliptic curve.  The fact that both the $\mathcal{N}=1^{*}$ and $\mathcal{N}=2^{*}$ theories are solved in the same way, using an elliptic curve, can probably be traced back to their common origin in $\mathcal{N}=4$ SYM.  As pointed out in \cite{dorey_exact_2002-1}, the fact that the matrix model can uncover this fact is remarkable.  

\section{$\mathcal{N}=4$ Super Yang-Mills Theory (SYM) in Four Dimensions}

Let the symmetry group of the theory be a compact Lie group $G$, and let $M$ be a four-dimensional spacetime manifold.  Take all fields to transform in the adjoint representation of G.  In addition, let $g_{\textnormal{YM}}$ denote the Yang-Mills coupling of the gauge theory.  This is \emph{not} to be confused with the coupling $g_{s}$ of a matrix model to come.

$\mathcal{N}=4$ SYM in four dimensions actually arises as a dimensional reduction of an even simpler theory: $\mathcal{N}=1$ SYM in ten dimensions.  This latter theory contains only vector multiplets $(A_{M}, \psi)$, where $A_{M}$ is a ten-dimensional gauge field with $M$ a ten-dimensional index, and $\psi$ is a 16 component Majorana-Weyl spinor.  We can dimensionally reduce this theory to four dimensions by regarding the latter six components of the gauge field as scalar fields which we call $\phi_{4}, \ldots, \phi_{9}$, while the first four components transform as a four-dimensional gauge field $A_{\mu}$.  In addition, there are right and left moving fermions.  The decomposition of the field content under this dimensional reduction is

\begin{equation}
\big(A_{M}, \psi \big) \longrightarrow  \big( A_{\mu}, \phi_{4}, \ldots, \phi_{9}, \psi^{L}, \psi^{R}, \chi^{L}, \chi^{R} \big).
\end{equation}
The resulting theory is known as $\mathcal{N}=4$ SYM in four dimensions.  This theory is superconformally invariant, even quantum mechanically.  This is surprising because, in general, Yang-Mills theories in four dimensions are \emph{classically} conformal, but this symmetry is broken by quantum anomalies.

Yang-Mills theories on four dimensional spacetimes are well known to have rich structure.  Specifically in four dimensions, one can add the following topological term to the Lagrangian of the theory,

\begin{equation}
\frac{\theta}{8 \pi^{2}} \int_{M} \textnormal{Tr} \, F \wedge F, 
\end{equation}
where $F$ is the curvature two-form associated to the gauge connection $A$, and $\theta$ is a coupling called the ``theta angle."  This quantity depends only on the topology of the principal $G$-bundle underlying the gauge theory, and is special to four-dimensional manifolds since $\textnormal{Tr} \, F \wedge F$ is a 4-form.  We can package the Yang-Mills coupling $g_{\textnormal{YM}}$ and the theta angle into the complexified gauge coupling,

\begin{equation} \label{eqn:gaugethcoup}
\tau_{0} = \frac{\theta}{2 \pi} + \frac{4 \pi i}{g_{\textnormal{YM}}^{2}}.
\end{equation}
It's crucial to note that $\tau_{0}$ is the coupling of the \emph{gauge theory}.  We will soon see a parameter $\tau$ arising from a matrix model, and the two are not to be identified a priori.  The fact that they are related is a very deep connection made by Dijkgraaf and Vafa, which we will come to later.  The conjectured S-duality asserts that $\mathcal{N}=4$ SYM in four dimensions is invariant under $\textnormal{SL}(2, \mathbb{Z})$ transformations of $\tau_{0}$.  These transformations are generated by $\tau_{0} \to \tau_{0}+1$ and $\tau_{0} \to -\tfrac{1}{\tau_{0}}$.  Thus, four-dimensional $\mathcal{N}=4$ SYM appears to be intrinsically connected to elliptic curves: (\ref{eqn:gaugethcoup}) essentially coordinatizes the moduli space of elliptic curves $\mathcal{M}_{1,1}$ with $\mathcal{N}=4$ SYM moduli.

\section*{The Mass Deformations of $\mathcal{N}=4$ SYM}

One often hears about re-writing the field content of some theory in the language of another theory.  This does not refer to any sort of deformation or dimensional reduction.  Rather, it's simply a repackaging of the field content of one theory, into a multiplet transforming properly under action by the symmetry group of another theory.  If we re-write the field content of $\mathcal{N}=4$ SYM in $\mathcal{N}=1$ language, we get a single $\mathcal{N}=1$ vector multiplet $(A_{\mu}, \lambda)$ and three chiral multiplets $\Phi_{i} = (Z_{i}, \psi_{i})$, for $i=1,2,3$.  Here, $Z_{i}$ are three complex scalar fields, $\lambda$ and $\psi_{i}$ are fermions, and $A_{\mu}$ is the four-dimensional gauge field.  The chiral multiplets interact via a cubic classical superpotential $\Phi_{1} [ \Phi_{2}, \Phi_{3}]$.  Notice that there are still 6 real scalar fields, one gauge field, and four fermions.  We have not gained or lost any data, we have only repackaged it.  We can just as easily rewrite the field content of $\mathcal{N}=4$ SYM in $\mathcal{N}=2$ language.  The data is packaged into a vector multiplet $(A_{\mu}, \Phi + i \Phi', \psi_{\alpha}^{1}, \psi_{\alpha}^{2})$, and two massless hypermultiplets $(\phi, \widetilde{\phi}, \chi_{\alpha}, \widetilde{\chi}_{\alpha})$.  Once we have repackaged the content of a theory, we can add certain mass terms to the action that break part of the original $\mathcal{N}=4$ supersymmetry.  We refer to these as massive deformations.  There are actually two unique massive deformations of $\mathcal{N}=4$ SYM which preserve $\mathcal{N}=1$ and $\mathcal{N}=2$ supersymmetry.  These are called the $\mathcal{N}=1^{*}$ and $\mathcal{N}=2^{*}$ theories, respectively.

\section{The $\mathcal{N}=1^{*}$ Theory and the Holomorphic Matrix Model}

In the introduction, I remarked that the $\mathcal{N}=1^{*}$ theory has a holomorphic subsector which consists of physical observables computable as derivatives of a particular function $W_{\textnormal{eff}}$ called the \emph{effective superpotential}.  The effective superpotential in a given classical vacuum must be a holomorphic function of the underlying parameters of the theory.  Thus, the most pressing matter in such theories becomes computing $W_{\textnormal{eff}}$.  If one can do so, then a whole class of physical observables are also easily obtainable.  In \cite{dijkgraaf_perturbative_2002}, Dijkgraaf and Vafa conjecture that one can compute $W_{\textnormal{eff}}$ for a given $\mathcal{N}=1^{*}$ vacuum by considering a particular holomorphic matrix model expanded around the vacuum under consideration.  It is this matrix model which will be studied.  I begin by defining $\mathcal{N}=1^{*}$ and studying the classical vacuum structure of the theory.  

\begin{defn}
The $\bm{\mathcal{N}=1^{*}}$ \textbf{theory} is realized as a massive deformation of $\mathcal{N}=4$ SYM by giving the same mass $m$ to each of the three chiral multiplets $\Phi_{i}$.  This adds quadratic mass terms to the classical superpotential:
\begin{equation} \label{eqn:classsuppot}
W_{\textbf{classical}}= \textnormal{Tr} \bigg(\Phi_{1} [\Phi_{2}, \Phi_{3}] + m\Phi_{1}^{2}+ m\Phi_{2}^{2}+ m\Phi_{3}^{2}\bigg).
\end{equation}
This mass deformation preserves $\mathcal{N}=1$ supersymmetry, which explains the name.  The original $\mathcal{N}=4$ SYM can be recovered by letting $m=0$.  
\end{defn}

\section*{The Classical Vacuum Structure}

Consider $\mathcal{N}=4$ SYM on $\mathbb{R}^{4}$ with gauge group $\textnormal{SU}(N)$, and complexified gauge coupling $\tau_{0}$.  Passing to the $\mathcal{N}=1^{*}$ massive deformation, we have a classical superpotential of the form (\ref{eqn:classsuppot}) where $m$ is the mass given to the chiral superfields.  We can make a convenient change of variables $\Phi^{+} \defeq \Phi_{1} + i \Phi_{2}$, $\Phi^{-} \defeq \Phi_{1} - i \Phi_{2}$, and $\Phi \defeq \Phi_{3}$.  In these variables, the classical superpotential takes the form

\begin{equation} \label{eqn:sup-pot}
W_{\textnormal{classical}}=\textnormal{Tr} \bigg(i \Phi [\Phi^{+}, \Phi^{-}] + m\Phi^{+} \Phi^{-} + m\Phi^{2}\bigg).
\end{equation}
The quantum field theory with action (\ref{eqn:sup-pot}) has a rich classical vacuum structure.  In \cite{dorey_massive_2002}, it is found that the classical vacua correspond to reducible or irreducible representations of $\textnormal{SU}(2)$.  In the representation theory of $\textnormal{SU}(2)$, one considers the $N \times N$ Hermitian matrix $J_{3}$, and the raising/lowering matrices $J^{\pm}$.  \emph{Classically}, $\Phi$ is identified with $iJ_{3}$, while $\Phi^{\pm}$ are related to $J^{\pm}$.  One can show that classically, the fields satisfy the equations,

\begin{equation}
\begin{split}
&[\Phi^{+}, \Phi^{-}] = 2 i m \Phi \\
&[\Phi \, , \Phi^{\pm}] = \pm \, i m \Phi^{\pm},
\end{split}
\end{equation}
which are simply anti-Hermitian versions of the familiar $\textnormal{SU}(2)$ commutation relations.  In general, we allow for reducible representations.  For all $p \vert N$ ($p$ dividing $N$) we can consider $N/p$ direct summands of a $p$-dimensional irreducible representation.  In \cite{dorey_massive_2002}, the authors find that each divisor $p$ of $N$ corresponds to a theory with reduced gauge group $\textnormal{SU}(N/p)$ and $N/p$ physically distinct vacua for this fixed $p$.  This means the total number of classical vacua is,

\begin{equation}
\sum_{p \vert N} N/p = \sum_{p \vert N} p.  
\end{equation}

\subsection*{The Confining and Higgs Vacua}

The \emph{confining vacuum} is defined by taking $p=1$, corresponding to $N$ indistinguishable vacua and the full $\textnormal{SU}(N)$ gauge group unbroken.  Recalling the representations consisted of $N/p$ copies of $p$-dimensional irreducible representations, in the confining vacuum, all fields can be taken to vanish,

\begin{equation}
\Phi = \Phi^{\pm} =0.
\end{equation}
Again, this holds only at a classical level; the fields don't vanish quantum mechanically.  The \emph{Higgs vacuum} is defined by taking $p=N$, which gives only a single vacuum with gauge group completely broken.  In this case, the representation consists simply of an irreducible $N$-dimensional representation.  The confining vacuum is particularly simple, and all other massive vacua will work in fundamentally the same way \cite{dorey_massive_2002}.  In this thesis, it will suffice to focus on the confining vacuum.

\section*{The Matrix Model}

The Dijkgraaf-Vafa conjecture \cite{dijkgraaf_perturbative_2002} is that to compute the effective superpotential in any vacuum, one should take the 't Hooft limit of a holomorphic three-matrix model expanded about the relevant vacuum.  This is particularly simple in the case of the confining vacuum since the classical fields vanish identically.  Therefore, we must consider the following holomorphic matrix model, which was originally studied in the thesis of Hoppe \cite{hoppe_quantum_1982},

\begin{equation}
\mathcal{Z}_{N}(m) = \int \mathcal{D} \Phi^{+} \mathcal{D} \Phi^{-} \mathcal{D} \Phi \, \,  e^{-\textnormal{Tr} \, W_{\textnormal{classical}}}.
\end{equation}
In the above partition function, $\Phi$ and $\Phi^{\pm}$ now represent $N \times N$ matrices independent of the classical solutions of the fields.  This is because we're looking at matrix fluctuations about the classical solutions.  We can scale all three matrices by $m$, and define the matrix model coupling $g_{s} \defeq 1/m^{3}$.  This produces a partition function with $g_{s}$ playing an analogous role to $\hbar$ in quantum mechanics

\begin{equation} \label{eqn:gsMM}
\mathcal{Z}_{N}(g_{s}) = \int \mathcal{D} \Phi^{+} \mathcal{D} \Phi^{-} \mathcal{D} \Phi \, \,  e^{-\frac{1}{g_{s}} \textnormal{Tr} \big( - \Phi[\Phi^{+}, \Phi^{-}] + \Phi^{+} \Phi^{-} + \Phi^{2}\big)}.
\end{equation}
The above three-matrix model is well-known in the literature, and was solved exactly by Kazakov, Kostov, and Nekrasov in \cite{kazakov_d-particles_1999}.  They showed that one can integrate over $\Phi^{\pm}$, leaving simply a one-matrix model,

\begin{equation} \label{eqn:partfuncc}
\mathcal{Z}_{N}(g_{s}) = \int \mathcal{D} \Phi \frac{e^{-\frac{1}{g_{s}} \textnormal{Tr} \Phi^{2}}}{\textnormal{det}(\textnormal{Adj}_{\Phi} + i)}.
\end{equation}
Here, $\textnormal{Adj}_{\Phi} = [\Phi, \, \cdot \,]$ is the adjoint action, and the commutator is the familiar commutator on matrices.  It is this holomorphic matrix model which I want to study in the 't Hooft limit.

\section*{The Solution of the Matrix Model in the 't Hooft Limit}

In \cite{kazakov_d-particles_1999} Kazakov, Kostov, and Nekrasov solve (\ref{eqn:partfuncc}) by passing to the eigenvalue representation, where the partition function reduces to,

\begin{equation} \label{eqn:eigpart}
\mathcal{Z}_{N}(g_{s}) = \int \prod_{I} d\lambda_{I} \prod_{I<J} \frac{(\lambda_{I}-\lambda_{J})^{2}}{(\lambda_{I} - \lambda_{J} + i)(\lambda_{I} - \lambda_{J} - i)}e^{-\frac{1}{g_{s}}\sum \lambda_{I}^{2}}.
\end{equation}
The equation of motion (the saddle-point equation) is computed to be,

\begin{equation} \label{eqn:eqnmot}
2 \lambda_{I} = g_{s} \sum_{I \neq J} \bigg( \frac{2}{\lambda_{I} - \lambda_{J}} - \frac{1}{\lambda_{I} - \lambda_{J} + i}- \frac{1}{\lambda_{I} - \lambda_{J} -i}\bigg).
\end{equation}
Taking the 't Hooft limit, we let $N \to \infty$ and $g_{s} \to 0$, holding fixed the 't Hooft coupling

\begin{equation}
S= g_{s}N.  
\end{equation}
In general, $S$ is complex-valued.  Notice that the above equation of motion has a familiar interpretation.  In the classical limit $g_{s} = 0$, the righthand side of (\ref{eqn:eqnmot}) vanishes, indicating that all $N$ eigenvalues are stacked at the critical point, which happens to be the origin in this case.  Taking the 't Hooft limit, one expects the eigenvalues to spread out into a continuous cut $[-\mu, \mu]$.  Since the matrix model is fundamentally holomorphic, the cut may be supported in the complex plane, but let us assume for now it is contained in the real axis.  As usual in matrix model technology, the resolvent is defined by

\begin{equation}
\omega(x) = \int_{-\mu}^{\mu} dy \, \frac{\rho(y)}{x-y},
\end{equation}
which is a holomorphic function on the complex $x$-plane, with a single branch cut corresponding to the matrix model eigenvalues.  The discontinuity of $\omega(x)$ across this cut gives the eigenvalue density

\begin{equation} 
\rho(x) = -\frac{1}{2 \pi i }\bigg(\omega(x + i \epsilon)- \omega(x - i \epsilon)\bigg).
\end{equation}
Consider a ``probe eigenvalue" located at some point $x$ in the complex plane.  This probe feels a force from the eigenvalues within the branch cut.  The force is given by,\footnote{The above $f$ is not to be confused with (\ref{eqn:fsum}).}

\begin{equation}
f(x) = 2x - S\bigg[ 2 \omega(x) - \omega(x + i) - \omega(x-i)\bigg],
\end{equation}
and clearly vanishes when $x$ lies in the branch cut, as can be seen from the equation of motion (\ref{eqn:eqnmot}).  The solution to the model relies on the definition of the function

\begin{equation}
G(x) \defeq x^{2} + iS\bigg[\omega(x + \tfrac{i}{2}) - \omega(x - \tfrac{i}{2})\bigg], 
\end{equation}
which is called the \emph{generalized resolvent} of $\mathcal{N}=1^{*}$.  It is straightforward to verify that $G(x)$ is related to the force through,

\begin{equation}
f(x) = -i \bigg[G(x + \tfrac{i}{2}) - G(x - \tfrac{i}{2})\bigg].
\end{equation}
Since the resolvent $\omega(x)$ has a single discontinuity along the branch cut, the function $G(x)$ is holomorphic with the exception of \emph{two} branch cuts.  These two branch cuts are the translates of the original branch cut by $\pm i/2$.  Keep in mind that there is only the single, original branch cut in which the eigenvalues live; these \emph{mirror branch cuts} are merely an artifact arising from $G(x)$.  For $x$ lying within the (real) cut, since $f(x)=0$, we see that for small $\epsilon$,

\begin{equation}
G(x + \tfrac{i}{2} \pm i \epsilon) = G(x- \tfrac{i}{2} \mp i \epsilon).
\end{equation}
This relation has the interpretation of a ``gluing condition" in the following sense: the top of the upper cut is glued to the bottom of the lower cut, and visa versa.  This condition implies that the generalized resolvent $G(x)$ is naturally identified as an elliptic function on an elliptic curve.  One can also show that $G(-x) = G(x)$ which constrains the geometry of the eigenvalue plane: the two branch cuts of $G(x)$ must be such that the entire eigenvalue plane is symmetric under inversion $x \to -x$.

We've now seen that the equation of motion of $\mathcal{N}=1^{*}$ on $\mathbb{R}^{4}$ in the 't Hooft limit has determined the generalized resolvent $G(x)$ as an elliptic function on an elliptic curve.  Later, the physical 't Hooft coupling $S$ of the matrix model will be related to the modular parameter $\tau$ of the elliptic curve.  The remaining solution of the model requires understanding how the elliptic curve is identified with the eigenvalue plane.  I will provide such a construction in the following chapter.

\section{The $\mathcal{N}=2^{*}$ Theory and its Critical Points}

Recall in $\mathcal{N}=2$ language, the field content of $\mathcal{N}=4$ SYM consists of a vector multiplet $(A_{\mu}, \Phi + i \Phi', \psi_{\alpha}^{1}, \psi_{\alpha}^{2})$ and two massless hypermultiplets $(\phi, \widetilde{\phi}, \chi_{\alpha}, \widetilde{\chi}_{\alpha})$.  The eigenvalues of the VEV of $\Phi$,

\begin{equation}
\langle \Phi \rangle = \textnormal{diag}(a_{1}, \ldots, a_{N}),
\end{equation}
are moduli on the Coulomb branch of $\mathcal{N}=2$.  We take the gauge group to be $G=\textnormal{SU}(N)$ such that all fields transform in the adjoint representation.  In addition, take the spacetime to be the four-sphere $S^{4}$ of radius $R$.

\begin{defn}
The $\bm{\mathcal{N}=2^{*}}$ \textbf{theory} is realized as a massive deformation of $\mathcal{N}=4$ SYM by giving equal masses $M \in \mathbb{R}$ to the two hypermultiplets.  This is the unique massive deformation which preserves $\mathcal{N}=2$ supersymmetry.  The original $\mathcal{N}=4$ SYM can be recovered by letting $M=0$.    
\end{defn}

In the introduction I mentioned Pestun's seminal method of localization which allows certain observables in $\mathcal{N}=2$ supersymmetric gauge theories to be solved exactly.  Here, I will be specifically interested in $\mathcal{N}=2^{*}$ compactified on the four-sphere $S^{4}$, in which case the partition function of the theory localizes to a finite-dimensional matrix model.  It is precisely this matrix model which will be studied in specific limits \cite{russo_massive_2013,russo_evidence_2013}.  I will be most interested in the decompactification limit which corresponds to sending the radius $R$ of the sphere to infinity.  In a loose sense, this unfolds $S^{4}$ into $\mathbb{R}^{4}$.  Recall it was on $\mathbb{R}^{4}$ that we studied $\mathcal{N}=1^{*}$, so the decompactification limit of $\mathcal{N}=2^{*}$ is where the two deformations may be related.  In addition, we take the large-$N$ 't Hooft limit, where $N$ corresponds to the size of the matrices in the matrix model.  In passing to the 't Hooft limit, both the weak and strong coupling regions of the matrix model will be considered.  Note that the coupling appearing in the matrix model partition function below (\ref{eqn:loccpart}) actually \emph{is} the gauge theory coupling,

\begin{equation}
\lambda = g_{\textnormal{YM}}^{2}N.
\end{equation}
This is to be contrasted with the previous model where the matrix model coupling $g_{s}$ has no a priori connection to the gauge theory.  In the 't Hooft and decompactification limits, there is a phase transition separating the weak and strong coupling regions.  In addition, as the 't Hooft coupling $\lambda$ increases ever more, we encounter an infinite sequence of phase transitions which accumulate at infinite coupling \cite{russo_massive_2013,russo_evidence_2013}.  There are no such phase transitions in $\mathcal{N}=1^{*}$ theory.

\section*{The Matrix Model}

As discussed above, we have compactified $\mathcal{N}=2^{*}$ theory to $S^{4}$ in order to make use of the localization results.  The localized partition function then takes the form of the following matrix model \cite{russo_massive_2013},

\begin{equation} \label{eqn:loccpart}
\mathcal{Z}_{N}(\lambda, M) = \int d^{N-1}a \prod_{I < J} \frac{(a_{I} - a_{J})^{2} H^{2}(a_{I}- a_{J})}{H(a_{I} - a_{J} -M) H(a_{I} - a_{J} + M)} e^{-\frac{8 \pi^{2} N}{\lambda} \sum a_{I}^{2}}.
\end{equation} 
I need to explain the components of this partition function.  The $a_{i}$ are the eigenvalues of the VEV of the adjoint scalar field $\Phi$ in the vector multiplet.  These are coordinates on the Coulomb branch moduli space of the $\mathcal{N}=2$ theory.  The 't Hooft coupling is denoted by $\lambda = g_{\textnormal{YM}}^{2} N$, and the function $H(x)$ is defined as,

\begin{equation}
H(x) \defeq \prod_{n=1}^{\infty} \bigg(1 + \frac{x^{2}}{n^{2}}\bigg)e^{-\frac{x^{2}}{n}}.
\end{equation}
In \cite{russo_massive_2013}, a factor of the Nekrasov instanton partition function $\mathcal{Z}_{\textnormal{inst}}$ is included, but I set it to 1 as it will not play a role in the present context.  In the 't Hooft limit, the path integral above is dominated by a saddle-point, which can be shown to satisfy the equation of motion

\begin{equation} \label{eqn:Zareqnmot}
\int_{-\mu}^{\mu} dy \rho(y) \bigg( \frac{1}{x-y} - \mathcal{K}(x-y) + \frac{1}{2} \mathcal{K}(x - y + M) + \frac{1}{2} \mathcal{K}(x -y -M)\bigg) = \frac{8 \pi^{2}}{\lambda} x,
\end{equation}
where the eigenvalue density

\begin{equation}
\rho(x) = \frac{1}{N} \sum_{I=1}^{N} \delta(x - a_{I}),
\end{equation}
becomes continuous in the 't Hooft limit, supported on the cut $[-\mu, \mu]$ and normalized to unity.  The function $\mathcal{K}(x)$ is essentially the logarithmic derivative of $H(x)$,

\begin{equation}
\mathcal{K}(x) \defeq - \frac{H'(x)}{H(x)} = 2x \sum_{n=1}^{\infty} \bigg(\frac{1}{n} - \frac{n}{n^{2} + x^{2}}\bigg).  
\end{equation}

\section*{The Decompactification Limit at Weak Coupling}

The decompactification limit corresponds to letting the radius $R$ of the sphere go to infinity.  In this case, the function $\mathcal{K}(x)$ can be approximated by its asymptotics at infinity \cite{russo_massive_2013},

\begin{equation}
\mathcal{K}(x) = x \log x^{2} + \mathcal{O}(x).
\end{equation}  
Using this approximation and differentiating the equations of motion (\ref{eqn:Zareqnmot}) once with respect to $x$ we get

\begin{equation} \label{eqn:reln}
\int_{- \mu}^{\mu} dy \rho(y) \log \bigg(\frac{M^{2}}{x^{2}} - 1\bigg)^{2} = \frac{16 \pi^{2}}{\lambda},
\end{equation}
which can be differentiated once again, resulting in

\begin{equation}  \label{eqn:saddpt}
\int_{-\mu}^{\mu} dy \rho(y) \bigg(\frac{2}{x-y} - \frac{1}{x-y + M} - \frac{1}{x-y-M}\bigg) = 0.  
\end{equation}
This is referred to as the equation of motion in the decompactification limit.  In the case of weak coupling, where $\mu \ll M$, the last two terms in equation (\ref{eqn:saddpt}) cancel, leaving the integral equation,

\begin{equation}
\int_{-\mu}^{\mu} dy \rho(y) \frac{1}{x-y} =0.
\end{equation}
This integral equation is well-known to have as its solution, the inverse square-root,

\begin{equation} \label{eqn:eigdens}
\rho(x) = \frac{1}{\pi \sqrt{\mu^{2} - x^{2}}}.
\end{equation}
 However, notice that this doesn't provide a relationship between the 't Hooft coupling $\lambda$ and the cut length $\mu$.  This relationship is provided by (\ref{eqn:reln}).  Once we have the expression for the eigenvalue density $\rho(x)$ in (\ref{eqn:eigdens}), we can plug it into (\ref{eqn:reln}), carry out the integration, and we find that

\begin{equation} \label{eqn:weakcutleng}
\mu = 2M e^{-\frac{4 \pi^{2}}{\lambda}}.
\end{equation}
This is how the cut-length, and 't Hooft coupling are related in the weak coupling region.  

The solution described above relied on an approximation in the $\mu \ll M$ limit.  However, at least in the weak coupling region, the model is actually amenable to an exact solution using elliptic curves in a nearly identical fashion to $\mathcal{N}=1^{*}$.  To pursue such a solution, an alternative interpretation of the equation of motion (\ref{eqn:saddpt}) is needed.  In the previous section we encountered the resolvent $\omega(x)$ and the eigenvalue density $\rho(x)$.  Using these same defintions, I now define the generalized resolvent of $\mathcal{N}=2^{*}$

\begin{equation}
\widetilde{G}(x) \defeq \omega\big( x + \tfrac{M}{2}\big) - \omega\big( x - \tfrac{M}{2}\big),
\end{equation}
which is a holomorphic function on the complex $x$-plane except now with two mirror branch cuts: the single branch cut of $\omega(x)$ has been translated by $\pm M/2$.  One can show by direct computation that the equation of motion (\ref{eqn:saddpt}) is equivalent to

\begin{equation} \label{eqn:glueingN2}
\widetilde{G}\big( x + \tfrac{M}{2} \pm i \epsilon\big) = \widetilde{G}\big( x - \tfrac{M}{2} \mp i \epsilon\big). 
\end{equation}
Just as in the previous section, this has the interpretation of a glueing condition: it prescribes that the top of one cut is to be identified with the bottom of the other, and visa versa.  Glueing the two cuts together, and adding a point at infinity, the eigenvalue plane is compactified into an elliptic curve.  We see that the equation of motion constrains the generalized resolvent $\widetilde{G}$ to be defined on an elliptic curve, exactly as in $\mathcal{N}=1^{*}$.

\section*{The Strong Coupling Region and Phase Transitions}

While the exact solution works for $\lambda$ small enough, it appears to break down when the 't Hooft coupling reaches a critical point of $\lambda_{c}^{(1)} \approx 35.4252$.  This critical point arises due to the two mirror branch cuts colliding at the origin.  In addition, there is an infinite sequence of phase transitions occurring at critical points $\lambda_{c}^{(2)} \approx 83$, $\lambda_{c}^{(3)} \approx 150$, and so on, which appear to be inaccessible to the exact solution, but have nevertheless been observed numerically \cite{russo_massive_2013,russo_evidence_2013}.  Russo and Zarembo interpret these critical points as the appearance of a new massless hypermultiplet in the spectrum.  They also give a concise expression for the cut length at the $n$-th critical point,

\begin{equation}  \label{eqn:cutcons}
\mu(\lambda_{c}^{(n)}) = \frac{n M}{2}.
\end{equation} 
Recalling the full cut is $[-\mu, \mu]$, quite simply the critical points of $\mathcal{N}=2^{*}$ arise each time the cut length coincides with an integer multiple of $M$.

\chapter{Embedding the Elliptic Curve into the Eigenvalue Plane}

In the last chapter, I reviewed two different physical theories whose equations of motion imply a deep connection to an elliptic curve.  This is another example of an algebraic variety emerging from a matrix model.  It turns out that the solution to both models will require explicitly constructing a map $x$ embedding the elliptic curve into the eigenvalue plane, compactified at infinity.  There will be virtually no physics in this portion of the solution.  It is a purely mathematical problem, independent of either the $\mathcal{N}=1^{*}$ or $\mathcal{N}=2^{*}$ theories.  The distinction between the two physical theories will be manifested later in computations through the choice of one of the two generalized resolvents.  

This chapter is motivated by a construction of Hollowood and Prem-Kumar \cite{hollowood_partition_2015}.  My original work consists of attempting to apply this method to an enlarged region of parameter values in  $\mathbb{H}$.  I find a large open set $\mathcal{H} \subset \mathbb{H}$ which I refer to as the \emph{moduli space}, such that for all $\tau \in \mathcal{H}$ and $M \in \mathbb{C}$ I can determine the configuration of the two mirror branch cuts in the eigenvalue plane as well as the unique non-trivial cycle on the elliptic curve mapping under $x$ into the cut.  The compliment $\mathbb{H} \setminus \mathcal{H}$ is a region where my construction degenerates and is somewhat shrouded in mystery; I can only speculate on its possible physical significance.  The region $\mathbb{H} \setminus \mathcal{H}$ has a boundary component I call the \emph{line of degeneration} such that approaching this line from within $\mathcal{H}$, the two mirror cuts are converging to an overlapping configuration (on the real axis, for $M \in \mathbb{R}$).  On the line of degeneration there are discrete points where the cut length is approaching integer multiples of $M$.  While completely arbitrary from a mathematical perspective, I will show in the next chapter that these points actually play a distinguished role in the $\mathcal{N}=1^{*}$ theory, and perhaps $\mathcal{N}=2^{*}$ as well.  

Finally, my geometrical construction in this chapter will allow for the computation of eigenvalue densities later.  I reviewed above the idea that the eigenvalue density in a matrix model is encoded as the discontinuity of the resolvent across the cut.  Given the cycle on the elliptic curve mapping into the cut (which I describe in this chapter), the discontinuity can be computed from the generalized resolvent restricted to the cycle.  This gives a simple prescription which allows for the computation of eigenvalue densities in either the $\mathcal{N}=1^{*}$ or $\mathcal{N}=2^{*}$ theories.

\section{The Construction}

Let $\omega_{1}$ and $\omega_{2}$ be two complex numbers such that their ratio is not real.  Given two such parameters, we can form the following lattice within the complex plane,

\begin{equation}
\Lambda = 2 \omega_{1} \mathbb{Z} \oplus 2 \omega_{2} \mathbb{Z}.
\end{equation}

\begin{defn}
The complex manifold $\mathbb{C}/\Lambda$ is called a \textbf{complex torus} with half-periods $\omega_{1}$ and $\omega_{2}$.  The \textbf{modular parameter} of the torus is defined to be
\begin{equation}
\tau = \frac{\omega_{2}}{\omega_{1}}.
\end{equation}
which is taken to live in the upper-half complex plane $\mathbb{H}$ without loss of generality.  An \textbf{elliptic curve}\footnote{The hyperelliptic curves defined in the first chapter are generalizations of elliptic curves.  An elliptic curve can be realized as a projective algebraic curve embedded in the projective plane $\mathbb{P}^{2}$.  Passing to an affine chart, the elliptic curve takes the form $y^{2} = F(x)$ where $F$ is a cubic polynomial with complex coefficients.} consists of an underlying complex torus whose points constitute an abelian group \cite{husemoller_elliptic_2010,koblitz_introduction_1993}.  Therefore, an elliptic curve is a complex torus along with the choice of a distinguished point to serve as the group identity.  
\end{defn}

Let $z$ be the coordinate on a complex torus, let $x$ be the coordinate on the eigenvalue plane, and denote the two branch cuts by $C^{+}$ and $C^{-}$.  We want to construct a map $x(z)$ which embeds the torus into the eigenvalue plane.  Note that we are using the same symbol $x$ to denote both the map and the coordinate on the plane.  What conditions must this map $x(z)$ satisfy?  From the geometry of the eigenvalue plane, we see that it must be a quasi-periodic function\footnote{A quasi-periodic function on $\mathbb{C}/ \Lambda$ is a function $f:\mathbb{C} \to \mathbb{C}$ such that $f(z + 2 \omega_{1}) = f(z) + C_{1}$ and $f(z + 2 \omega_{2}) = f(z) + C_{2}$, where $C_{1}$ and $C_{2}$ are complex numbers.} on $\mathbb{C}/\Lambda$.  That is to say,

\begin{equation} \label{eqn:z(u)}
x(z+2\omega_{1})=x(z), \,\,\,\,\,\,\,\,\,\,\,\,\,\,\,\, x(z+2 \omega_{2})=x(z)+M,
\end{equation}
where $\pm M/2$ are precisely the midpoints of the two branch cuts in the $x$ plane.  These conditions uniquely determine $x(z)$ in terms of the Weierstrass $\zeta$-function,

\begin{equation} \label{eqn:zeta}
x(z)= iM \frac{\omega_{1}}{\pi}\bigg(\zeta(z)-\frac{\zeta(\omega_{1})}{\omega_{1}}z\bigg).
\end{equation}
This can be easily shown to obey the conditions in (\ref{eqn:z(u)}) by noting that the Weierstrass $\zeta$-function is quasi-periodic along both periods,

\begin{equation}
\zeta(z+2 \omega_{1,2})=\zeta(z)+2 \zeta(\omega_{1,2}).
\end{equation}
Without loss of generality, we can gauge fix one of the half-periods to be real.  For convenience, we make the following choice,

\begin{equation} 
\omega_{1}= \frac{\pi}{2}, \,\,\,\,\,\,\,\,\,\, \omega_{2}= \frac{\pi \tau}{2}.
\end{equation}
With this choice, the embedding becomes

\begin{equation} \label{eqn:emb}
x(z)=\frac{iM}{2}\bigg[\zeta(z)-\frac{1}{3}E_{2}(\tau) z \bigg].
\end{equation}
We observe that the points on the torus $z= \pm \omega_{2}$ map into the midpoints of the cuts,\footnote{This statement can be easily verified noting that $\zeta$ is an odd function satisfying $\omega_{2} \zeta(\omega_{1})-\omega_{1} \zeta(\omega_{2}) = i \pi/2$ as well as the relationship with the Eisenstein series $12 \omega_{1} \zeta(\omega_{1})=\pi^{2} E_{2}(\tau)$.  See Appendix A of \cite{dorey_exact_2002-1}.}

\begin{equation}
x(\pm \omega_{2})=\pm \frac{M}{2}.  
\end{equation}
The Weierstrass $\zeta$-function has a simple pole at $z=0$ on the torus.  It follows that $x(z)$ itself has a simple pole at $z=0$,

\begin{equation}
x(z)\big|_{z \to 0} \approx \frac{iM}{2} \frac{1}{z} + \ldots,
\end{equation}
and we conclude that $z=0$ on the torus maps to the point at infinity on the eigenvalue plane.


Recall that an elliptic curve is simply a complex torus with a distinguished point.  The above construction has produced such a distinguished point.  As $\tau$ varies, generic points in $\mathbb{C}/ \Lambda$ take various values under $x$.  However, zero is \emph{always} mapped to the point at infinity and as such, we have a distinguished point in a natural way.  It follows that this is a correspondence to elliptic curves, not merely complex tori.  
\vskip3ex

\begin{figure}
\begin{center}
\includegraphics[width=9cm]{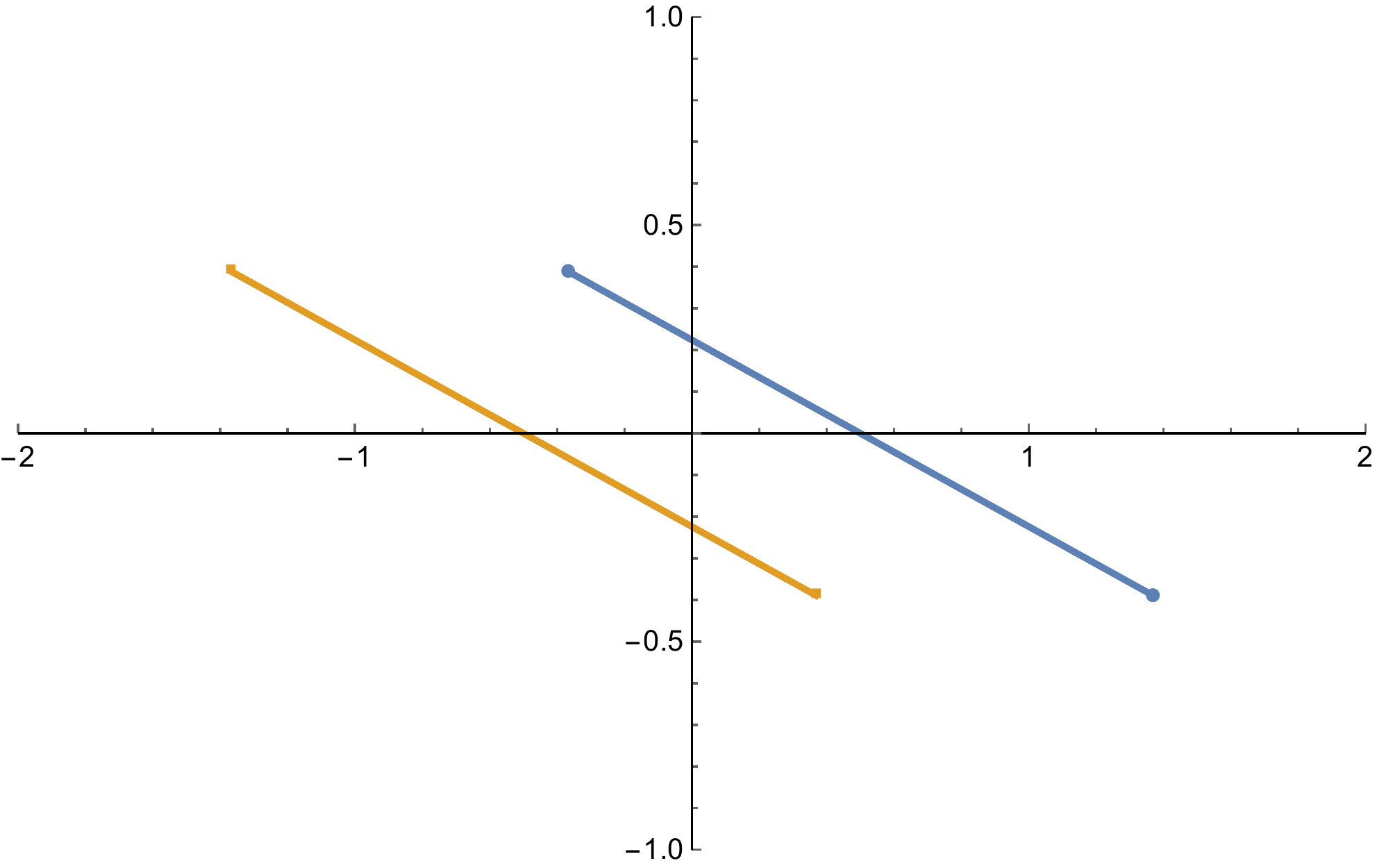}
\caption{The configuration of the eigenvalue plane for $\tau = \frac{1}{4} + \frac{1}{4} i$ and $M=1$.  Notice that the midpoints of the cuts are $\pm \tfrac{M}{2}$.  As well, note the symmetry of the eigenvalue plane. \label{fig:cutgeo1}}
\end{center}
\end{figure}

Since it is a torus, an elliptic curve has two linearly independent homology cycles, usually called $A$- and $B$-cycles.  Under $x$, the $A$-cycles will map into the eigenvalue plane such that their image encircles one of the branch cuts, while the image under $x$ of the $B$-cycles will be a path connecting the two branch cuts.  There must exist two special $A$-cycles, which map into the two branch cuts, respectively.  More precisely, the cycles double cover the branch cuts; one should imagine the images as ``infinitely tightly" encircling them.  Since the cuts are glued together, these two cycles are really the same, though they will appear translated by $2\omega_{2}$ on the fundamental domain of the elliptic curve.  Let us call these two special $A$-cycles $\mathcal{C}^{+}$ and $\mathcal{C}^{-}$, not to be confused with the branch cuts $C^{+}$ and $C^{-}$.  We must have $\omega_{2} \in \mathcal{C}^{+}$ and $-\omega_{2} \in \mathcal{C}^{-}$, which is to say that $\mathcal{C}^{+}$ maps into the cut with midpoint $\tfrac{M}{2}$ while $\mathcal{C}^{-}$ maps into the cut with midpoint $-\tfrac{M}{2}$.  These special $A$-cycles are related to the branch cuts through the embedding as

\begin{equation}
x ({\mathcal{C}^{+}}) = C^{+}, \,\,\,\,\,\,\,\,\, x({\mathcal{C}^{-}})= C^{-}.
\end{equation}
How do we find $\mathcal{C}^{+}$ and $\mathcal{C}^{-}$?  The following method is provided in \cite{hollowood_partition_2015}.  Choose a fixed value of the modular parameter $\tau \in \mathbb{H}$.  This fixes an elliptic curve with half-periods $\omega_{1}$ and $\omega_{2}$, as described above.  Since the two cycles $\mathcal{C}^{+}$ and $\mathcal{C}^{-}$ map directly into the cuts, and cover them twice, at the endpoints of the branch cuts, the derivative of $x(z)$ must vanish.  Thus, the condition that $x'(z)=0$ determines the points in the fundamental domain of the elliptic curve, which map into the branch points.  From (\ref{eqn:emb}), using $\zeta'(z) = -\wp(z)$, we find the derivative of the embedding to be,

\begin{equation}
x'(z)=-\frac{iM}{2}\big[\wp(z)+\tfrac{1}{3}E_{2}(\tau)\big].
\end{equation} 
Setting this to zero, and noting that $E_{2}(\tau)$ is simply a complex number for fixed $\tau$, we get a transcendental equation

\begin{equation} \label{eqn:transeq}
\wp(z) = -\frac{1}{3}E_{2}(\tau).  
\end{equation}
The Weierstrass $\wp$-function $\wp(z)$ takes every value in $\mathbb{C}$ exactly twice as we range over the fundamental domain of the elliptic curve.  Therefore, the above equation has exactly two solutions, call them $z_{1}$ and $z_{2}$.  It is precisely these two points which map to the endpoints of one of the branch cuts.  In fact, translating these two points by $2\omega_{2}$ will give two more points satisfying $x'(z)=0$, and these will map to the endpoints of the other branch cut.  The four points described here, have images in the eigenvalue plane such that their values all add up to zero.  In other words, the branch points satisfy the symmetry $x \to -x$ of the eigenvalue plane (Figure \ref{fig:cutgeo1}).

What I have discussed thus far has been entirely independent of the choice of branch cut.  In principle, there's some array of possible choices of the (\emph{real}) cut which maintains the symmetry of the eigenvalue plane under inversion, when the cut is translated by $\pm \tfrac{M}{2}$.  For my purposes, it will suffice to take the branch cuts to be straight line segments.  

\begin{figure}
\begin{center}
\begin{overpic}[width=10cm,height=11cm]{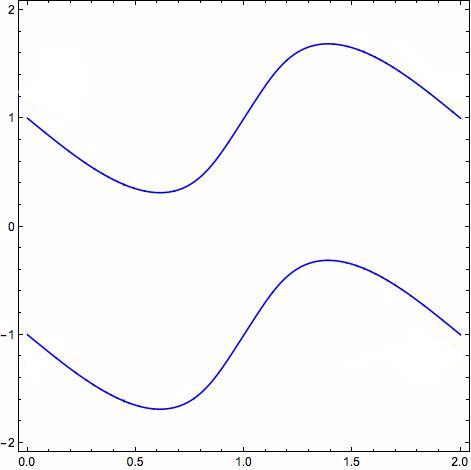}
 \put (15,75) {\huge$\displaystyle\mathcal{C}^{+}$}
 \put (74,25) {\huge$\displaystyle\mathcal{C}^{-}$}
\end{overpic}
\caption{The fundamental domain of the elliptic curve defined as the region between the two cycles $\mathcal{C}^{+}$ and $\mathcal{C}^{-}$, shown here for $\tau = \frac{1}{4} + \frac{1}{4} i$.  In this figure I use coordinates $\omega_{1}x + \omega_{2}y$ with $0 \leq x \leq 2$ and $-2 \leq y \leq 2$.  The upper contour is precisely $\mathcal{C}^{+}$ which maps into $C^{+}$ and the lower contour is $\mathcal{C}^{-}$ mapping into $C^{-}$.  Notice that the fundamental domain contains the points $\omega_{2}$, $-\omega_{2}$, $\omega_{2} + 2 \omega_{1}$, and $-\omega_{2} + 2 \omega_{1}$. \label{fig:funddom1}}
\end{center}
\end{figure}


Having decided on the shape of the branch cuts, we will be able to find the special cycles which map into the cuts.  The cycle $\mathcal{C}^{+}$ must contain the two points $z_{1}$ and $z_{2}$ mapping to the branch cuts, as well as $\omega_{2}$ and $\omega_{2} + 2 \omega_{1}$.  Translating $\mathcal{C}^{+}$ down by $2 \omega_{2}$ gives us $\mathcal{C}^{-}$.  Clearly, $\mathcal{C}^{-}$ must contain the translates of $z_{1}$ and $z_{2}$ as well as $-\omega_{2}$ and $-\omega_{2} + 2\omega_{1}$.  These two cycles \emph{define} the fundamental domain of the elliptic curve (Figure \ref{fig:funddom1}), meaning we choose the fundamental domain to be the region between $\mathcal{C}^{+}$ and $\mathcal{C}^{-}$.

\section{A Summary of the Method}

\begin{enumerate}
\vskip1ex

\item Choose $\tau \in \mathbb{H}$ and $M \in \mathbb{C}$.  If $\tau$ has too small an imaginary part, the model may have degenerated.  This ``region of degeneration" will be described shortly.  
\vskip1ex

\item By solving (\ref{eqn:transeq}), this $\tau$ determines four points $z_{1}, z_{2}, z_{1} - 2 \omega_{2}, z_{2} - 2\omega_{2}$ which map under $x(z)$ into the branch points on the eigenvalue plane.  Connect them pairwise with straight line segments in the unique way such that the midpoints of the two lines are $\pm \tfrac{M}{2}$.  Note that the resulting eigenvalue plane is invariant under $x \to - x$ (Figure \ref{fig:cutgeo1}).  
\vskip1ex

\item Find the $A$-cycle $\mathcal{C}^{+}$ on the elliptic curve passing through the points $\omega_{2}, z_{1}, z_{2},\\ \omega_{2} + 2\omega_{1}$ on the fundamental domain, which maps into the cut whose midpoint is $\tfrac{M}{2}$.  There will exist another $A$-cycle $\mathcal{C}^{-}$ translated down precisely by $2 \omega_{2}$ mapping into the other cut.  
\vskip1ex

\item If anything goes wrong so far, one has either chosen $\tau$ with too small an imaginary part, or the points $z_{1}$ and $z_{2}$ have not been correctly identified.  
\vskip1ex

\item Once the cycles have been found correctly, the fundamental domain of the torus is defined to be the region between the two $A$-cycles (Figure \ref{fig:funddom1}).  This fundamental domain contains a few special points: the origin $0$ of the elliptic curve maps into the point at infinity of eigenvalue plane, and the points $\pm \omega_{2}$ map into the midpoints $\pm \tfrac{M}{2}$ of the cuts.  
\vskip1ex

\item A key object in my prescription is the image $x(\mathcal{C}^{+}) = C^{+}$ of the $A$-cycle $\mathcal{C}^{+}$.  Obviously, one can plot the real or imaginary parts of this image as a function of a parameter on $\mathcal{C}^{+}$.  Either will suffice to compute which two points in $\mathcal{C}^{+}$ map to any given point in $C^{+}$.  This will allow for the computation of eigenvalue densities in a later chapter.  
\end{enumerate}

\section{Schematic Description of the Moduli Space of the Theory}

\begin{figure}
\begin{center}
\includegraphics[width=12.5cm]{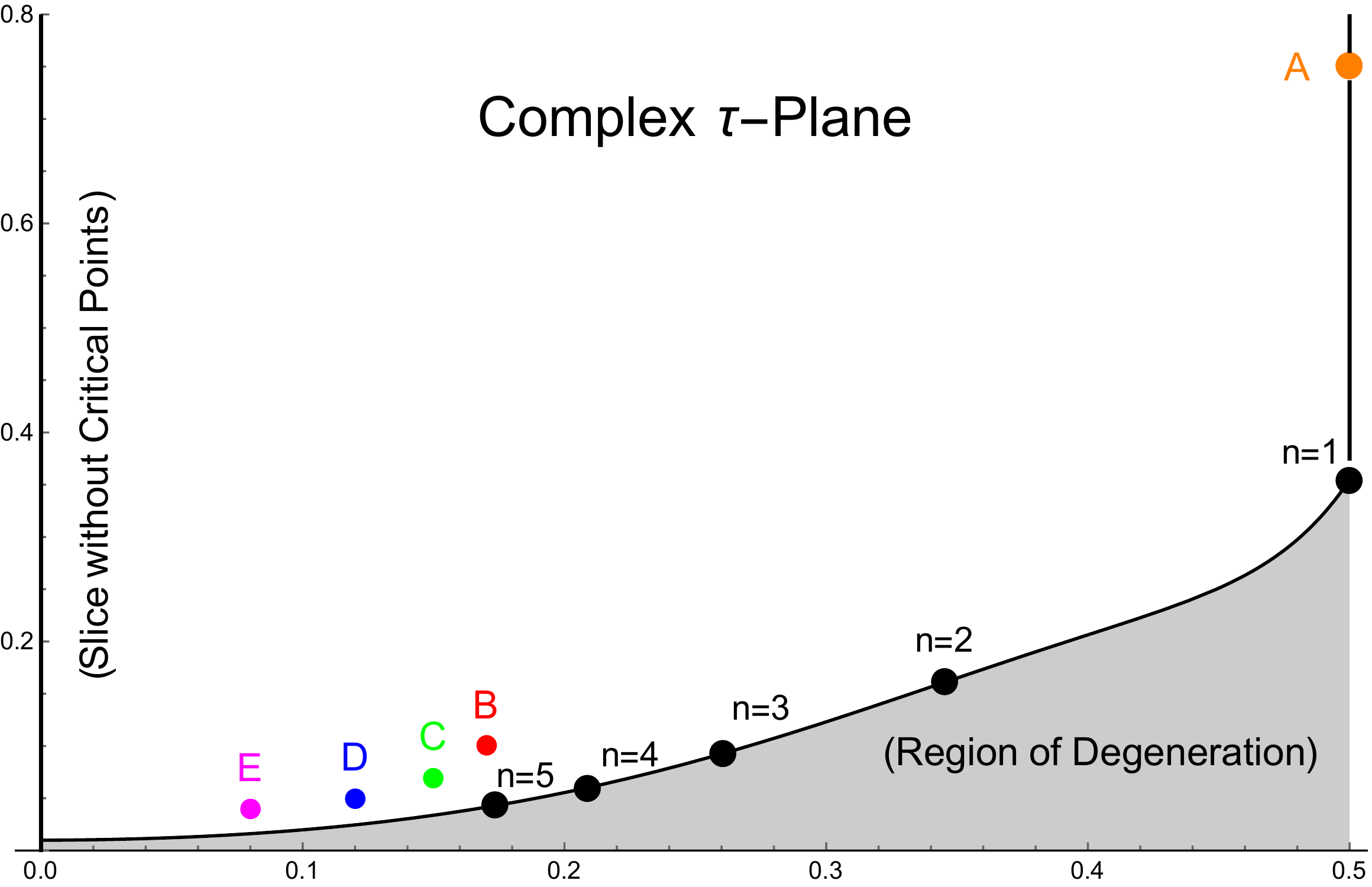}
\caption{A schematic diagram depicting (half of) the moduli space $\mathcal{H}$.  At the point labelled $n=1$, the two mirror branch cuts collide at the origin and each have length $|M|$.  At the points $n=2, 3, 4, 5, \ldots$ the mirror branch cuts are approaching overlap, each with length $n|M|$, respectively.  The points $A-E$ are labeled for future reference to indicate locations in the moduli space where eigenvalue densities are computed. \label{fig:modspplot}}
\end{center}
\end{figure}

Above, I provide a method such that for $\tau$ in some allowed region of the upper-half plane, one can construct the two branch cuts on the eigenvalue plane, as well as the cycles on the elliptic curve mapping into the cuts.  I now want to schematically describe this region of allowed $\tau \in \mathbb{H}$, which will be referred to as the \emph{moduli space}.  The relevant equations above are explicitly invariant under $\tau \to \tau + 1$, so we may restrict to the strip $\{0 \leq \text{Re}(\tau) < 1 \} \subseteq \mathbb{H}$.  Moreover, there is symmetry observed with respect to reflection across the line $\text{Re}(\tau)=\frac{1}{2}$, so it suffices to consider only $0 \leq \text{Re}(\tau) \leq \frac{1}{2}$.  For all $0 < \text{Re}(\tau) \leq \frac{1}{2}$, there exists a small enough $\text{Im}(\tau)$ such that the branch cuts collide on top of one another on the real axis (see for instance, Figure \ref{fig:CriticalPoints1}).  This defines a one-dimensional path in $\mathbb{H}$ which I call the \emph{line of degeneration}.\footnote{I was not able to find an explicit form of this line of degeneration.}  For yet smaller $\text{Im}(\tau)$, the model has completely degenerated.  This is referred to as the \emph{region of degeneration}.  The portion of the infinite strip $\{0 \leq \text{Re}(\tau) < 1 \} \subseteq \mathbb{H}$ \emph{above} the line of degeneration is the moduli space of the theory, which I denote by $\mathcal{H}$ (Figure \ref{fig:modspplot}).  Note that in the figure, I only plot half of the moduli space since $\mathcal{H}$ behaves symmetrically with respect to reflection about the line $\text{Re}(\tau) = \frac{1}{2}$.

Let me briefly summarize some of the schematic features of $\mathcal{H}$ which will play a role in the following chapters, as well as distinguished slices:

\begin{itemize}
\item It is natural to call $\mathcal{H}$ the matrix model moduli space: every branch cut configuration is realized once, and only once, in $\mathcal{H}$.  

\item Varying $M \in \mathbb{C}$ does not change the relative position of the two mirror cuts; it merely changes the midpoints and rotates both cuts by $\text{Arg}(M)$.  With this in mind, for simplicity I will usually choose $M \in \mathbb{R}$.  

\item For $M \in \mathbb{R}$, I refer to $\text{Re}(\tau) =0$ as the \emph{anti-Hermitian slice}, as both branch cuts lie parallel to the imaginary axis in the eigenvalue plane.  As such, they can freely enlarge without colliding.

\item Again for $M \in \mathbb{R}$, the \emph{Hermitian slice} is $\text{Re}(\tau) =\frac{1}{2}$ as the branch cuts are supported symmetrically in the real axis.  However, we can see that the cuts will collide at the critical point, labeled $n=1$.

\item Both the Hermitian and anti-Hermitian slices intersect at infinity in the upper-half plane.  At this point, the cut length vanishes.  Emerging from the point at infinity along a particular slice through $\mathcal{H}$ corresponds to the branch points spreading out at a particular angle in the eigenvalue plane.  This holds only in a neighborhood of infinity; I will show in the following chapter that there is a non-trivial relationship between the branch cut configuration and the position within $\mathcal{H}$.  

\item There is a region within $\mathcal{H}$ where the two mirror cuts can be made \emph{arbitrarily} long and approaching an overlapping configuration.  For an arbitrarily small $\text{Re}(\tau)$, there will be a correspondingly small $\text{Im}(\tau)$ giving rise to such a configuration.  One question pursued in the following chapter is, assuming there is a Hermitian matrix model whose weak coupling region naturally embeds along the Hermitian slice, \emph{is the strong coupling region embedded along the line of degeneration?}  Does approaching this line from within $\mathcal{H}$ shed light on the strong coupling behavior of the theory?    

\item Along the line of degeneration, there exists an infinite collection of points $n=1,2,\ldots$ where the two cuts are approaching degeneration in the real axis each with length $n|M|$. The first five of these are labeled $n=1, \ldots, 5$ in Figure \ref{fig:modspplot}.  Though seemingly arbitrary at this point, these points actually play a special role in $\mathcal{N}=1^{*}$, and possibly $\mathcal{N}=2^{*}$ as well.    
\end{itemize}

Having completed the construction, I want to re-emphasize that this chapter has been independent of physics.  Moreover, the ``branch cuts" appearing here need not even be the cuts of a matrix model in the 't Hooft limit; this was a purely mathematical pursuit.  \emph{I now want to realize my construction as the underlying structural backbone of a physical holomorphic matrix model.}  One could restrict this model to either the Hermitian or anti-Hermitian slices, but doing so would veil the bulk of $\mathcal{H}$ where I have noted potentially interesting phenomena.  My goal is to now reconnect with the $\mathcal{N}=1^{*}$ and $\mathcal{N}=2^{*}$ theories and study the physical significance of these phenomena.

\chapter{The Quasi-Modular Coupling $S(\tau)$}

The construction in the previous chapter was independent of any physical theory.  In a sense, one should think of it as the geometrical backbone of any matrix model whose equation of motion identifies the eigenvalue plane with an elliptic curve.  The physical content of a particular theory is then encoded into the generalized resolvent.  It should be expected that all physical quantities (eigenvalue densities, physical couplings, etc.) will be computed using both the geometrical construction of the last chapter, as well as the generalized resolvents.  Eigenvalue densities will occupy the following chapter but presently, I want to describe a computation of the matrix model 't Hooft coupling in the $\mathcal{N}=1^{*}$ theory, which appeared originally in \cite{dijkgraaf_perturbative_2002, dorey_exact_2002-1}.  In this context, the physical parameters of the theory appear naturally in terms of the modular parameter $\tau$ of the elliptic curve, perhaps for all $\tau \in \mathcal{H}$, or perhaps constrained to a particular slice.  The only physical parameters in the $\mathcal{N}=1^{*}$ matrix model are the mass $M$ as well as the 't Hooft coupling $S$.  The mass is more or less a bystander, but the 't Hooft coupling emerges as a holomorphic function $S(\tau)$ on the upper-half plane, and a quasi-modular form with respect to $\text{SL}(2, \mathbb{Z})$.  

The Fourier expansion of $S(\tau)$ is really an expansion at infinity in the upper-half plane, which I will show to be the weak coupling region of the theory.  This point at infinity is the unique zero of $S(\tau)$ in the open set $\mathcal{H}$ and is also the unique point where the matrix model cut length vanishes.  I will argue here that in a neighborhood of infinity, $S(\tau)$ determines the exact branch cut configuration of the matrix model.\footnote{More specifically, $S(\tau)$ determines the function $Q(\tau)$ I define in (\ref{eqn:MMConfig}) which in turn, determines the weak coupling configuration of the matrix model.}  This breaks down at higher coupling, but I hope to show that $S(\tau)$ actually encodes non-perturbative data about \emph{both} the $\mathcal{N}=1^{*}$ and $\mathcal{N}=2^{*}$ models.  A collection of extrema of $S(\tau)$ fall on the line of degeneration.  Approaching these points, the matrix model cut length converges to integral multiples of $M$.  This appears to hold at arbitrarily high coupling.  One of the extrema of $S(\tau)$ lies on the Hermitian slice $\text{Re}(\tau)=\frac{1}{2}$ and remarkably, this is \emph{exactly} the first critical point of $\mathcal{N}=2^{*}$.  It is only by considering the full holomorphic nature of $\mathcal{N}=1^{*}$ that one is able to see the extra data it encodes, which as far as I am aware, is not explicit in the literature.

\section{The 't Hooft Coupling of $\mathcal{N}=1^{*}$}

In the original papers on the $\mathcal{N}=1^{*}$ matrix model \cite{dijkgraaf_perturbative_2002, dorey_exact_2002-1,dorey_massive_2002}, the authors take $M=i$.  With this choice, equation (\ref{eqn:emb}) becomes

\begin{equation}
x(z)=-\frac{1}{2}\bigg[\zeta(z)-\frac{1}{3}E_{2}(\tau) z \bigg].
\end{equation}
This agrees precisely with equation (2.14) in \cite{dorey_exact_2002-1}, noting that $\zeta(\omega_{1}) \omega_{1} = \tfrac{\pi^{2}}{12} E_{2}(\tau)$ and recalling the choice $\omega_{1} = \tfrac{\pi}{2}$.  Also in \cite{dorey_exact_2002-1} Dorey et. al. provide an expression for the generalized resolvent.  One can easily show that their equation (2.15), using the choice of $\omega_{1}$, is given by

\begin{equation} \label{eqn:N1starresolve}
G(x(z)) = \frac{M^{2}}{4} \bigg[\wp(z) - \frac{2}{3} E_{2}(\tau)\bigg].
\end{equation}
The matrix model 't Hooft coupling $S$ is related to the generalized resolvent through a contour integral on the eigenvalue plane, surrounding the upper branch cut,

\begin{equation} \label{eqn:N=1*thooft}
S = g_{s} N = \frac{1}{2 \pi i} \oint_{C^{+}} G(x) dx = \frac{-1}{2\pi} \oint_{\mathcal{C}^{+}} G(x(z)) \frac{dx}{dz} dz,
\end{equation}
where the final equality follows by changing variables and integrating over the corresponding $A$-cycle on the torus mapping into the upper cut under the embedding.  It's relatively straightforward to carry out the integration on the torus.  We begin by noting that, since $\oint_{C^{+}} dx =0$,

\begin{equation} \label{eqn:acycvan}
0 = \oint_{C^{+}} dx = \oint_{\mathcal{C}^{+}} \frac{dx}{dz} dz = \frac{1}{2} \oint_{\mathcal{C}^{+}} \bigg[\wp(z) + \frac{1}{3}E_{2}(\tau)\bigg]dz.
\end{equation}
In the final equality, I have used that $\zeta'(z) = -\wp(z)$.  Equation (\ref{eqn:acycvan}) immediately implies the relationship,

\begin{equation}
\oint_{\mathcal{C}^{+}} \wp(z) dz =  -\frac{\pi}{3}E_{2}(\tau),
\end{equation}
where I used $\oint_{\mathcal{C}^{+}} dz = 2\omega_{1}$.  Similarly, one can show that

\begin{equation}
\oint_{\mathcal{C}^{+}} \wp^{2}(z) dz = \frac{\pi}{9} E_{4}(\tau).
\end{equation}
The reader should be warned that in the above formulae I have used specifically the choice of $\omega_{1} = \tfrac{\pi}{2}$; these formulae may be seen in the literature with explicit dependence on $\omega_{1}$, or may differ by a minus sign thanks to a particular choice.  Using the above results, the 't Hooft coupling $S$ can finally be computed in terms of the modular parameter $\tau$ on the elliptic curve:

\begin{equation}
\begin{split}
S & =  -\frac{1}{2 \pi} \oint_{\mathcal{C}^{+}} G(x(z))\frac{dx}{dz} dz  \\
&= -\frac{1}{2 \pi} \oint_{\mathcal{C}^{+}} \frac{1}{8} \wp(z) \bigg[\wp(z) + \frac{1}{3}E_{2}(\tau)\bigg]dz \\
&= - \frac{1}{16 \pi} \oint_{\mathcal{C}^{+}} \wp^{2}(z) dz - \frac{E_{2}(\tau)}{48 \pi} \oint_{\mathcal{C}^{+}} \wp(z) dz \\
&= \frac{1}{144}\big(E_{2}(\tau)^{2} - E_{4}(\tau)\big).
\end{split}
\end{equation}
This agrees precisely with \cite{dijkgraaf_perturbative_2002,dorey_exact_2002-1}.  
\vskip2ex
\begin{equation}
\boxed{S(\tau) = \frac{1}{144}\big(E_{2}(\tau)^{2} - E_{4}(\tau)\big)}
\end{equation}
\vskip2ex
\noindent Since $S$ is the matrix model 't Hooft coupling, one should think of $S(\tau)$ as a ``change of variables" from the modular parameter $\tau$ to the physical coupling $S$.  I would like to record a few observations about $S(\tau)$: 

\begin{itemize}

\item $S(\tau)$ is a quantity belonging specifically to $\mathcal{N}=1^{*}$ theory, as is clear in the computation where the generalized resolvent of $\mathcal{N}=1^{*}$ was chosen.

\item This function is holomorphic on the upper-half plane $\mathbb{H}$.  Using the symmetries of the Eisenstein series, it's clear that $S(\tau + 1) = S(\tau)$.  Due to this symmetry, $S$ descends to a holomorphic function on an infinite strip of width one.  In Figure \ref{fig:ModuliSpace2} I plot the \emph{argument} of $S(\tau)$ over this infinite strip.  The left half of Figure \ref{fig:ModuliSpace2} should be compared to the schematic diagram in Figure \ref{fig:modspplot}.

\item Mathematically, $S(\tau)$ is a \emph{quasi-modular cusp form of weight four}.  It is called a \emph{cusp form} because it vanishes at the point at infinity in the upper-half plane.

\begin{figure}
\begin{center}
\includegraphics[width=12cm]{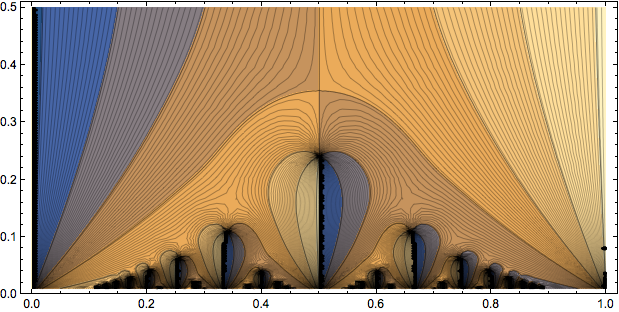}
\caption{A plot of the argument of $S(\tau)$. \label{fig:ModuliSpace2}}
\end{center}
\end{figure}

\item In Figure \ref{fig:ModuliSpace2} we can see a sequence of ``saddle-points" $\tau_{c}^{(1)}, \tau_{c}^{(2)}, \ldots$ with increasingly small real and imaginary parts which appear to accumulate near the origin.  One can check that these are the extrema of the 't Hooft coupling $S(\tau)$,

\begin{equation}
\frac{dS}{d\tau}(\tau_{c}^{(n)}) = 0 \,\,\,\,\,\,\,\, (n \geq 1).
\end{equation}
Applying the methods from the previous chapter, we see that these extrema of $S$ fall on the line of degeneration of my model and are therefore, technically inaccessible.  However, we may approach them from within $\mathcal{H}$ and doing so, leads to a non-obvious result:  for all $M \in \mathbb{C}$, approaching $\tau_{c}^{(n)}$ from within $\mathcal{H}$, we find that the resulting branch cut configuration consists of the two mirror cuts converging to an overlapping configuration, each of length $n|M|$ (Figure \ref{fig:CriticalPoints1}).

\begin{figure}
\includegraphics[width=13cm,height=12cm]{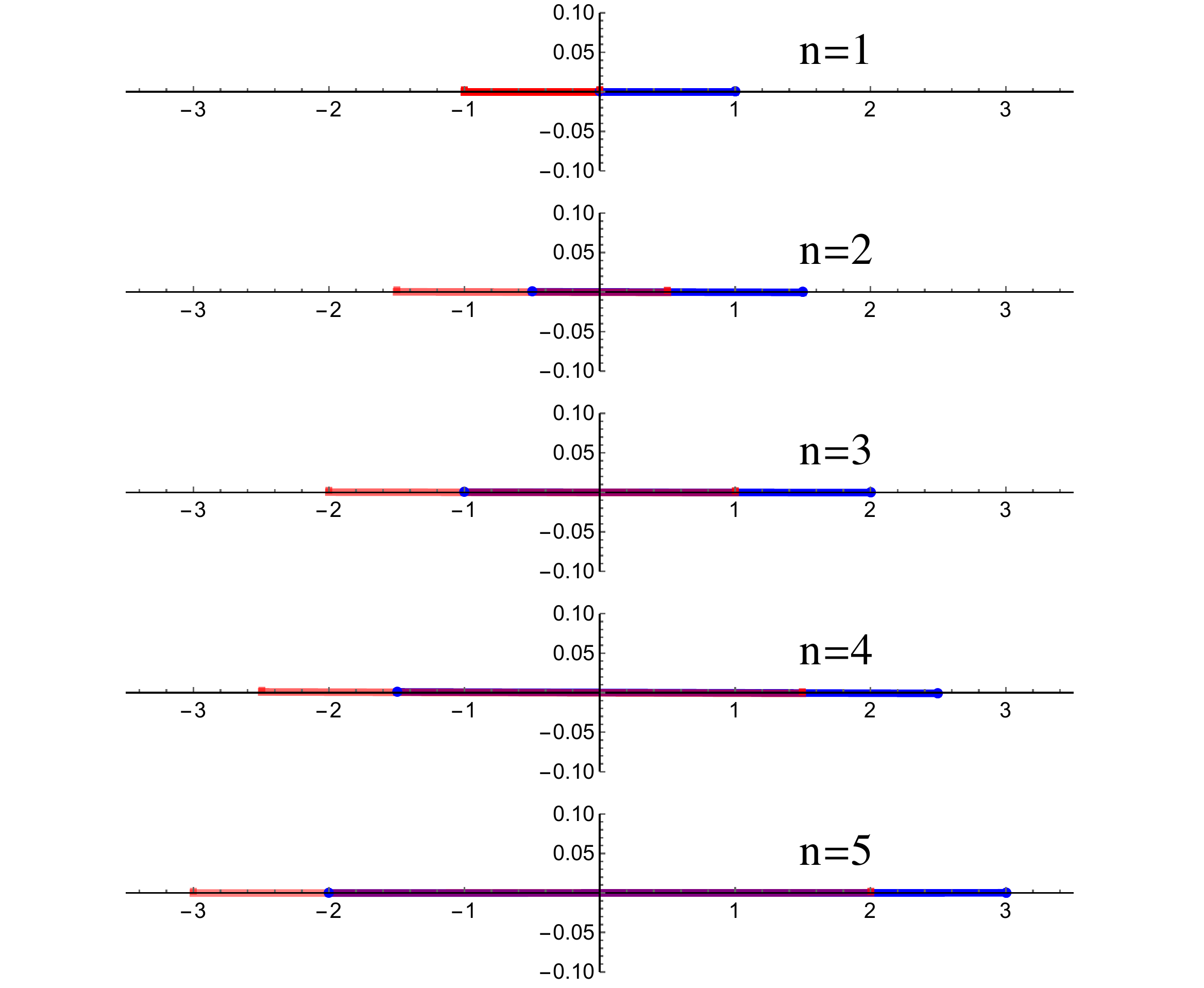}
\caption{The configuration of the eigenvalue plane at the first five critical points of $S(\tau)$ with $M=1$.  It's important to note that the cuts don't actually overlap here, rather they're both converging to the real axis, as the model nears degeneration. \label{fig:CriticalPoints1}}
\end{figure}

\item I have already noted that $S$ vanishes at infinity in the upper-half plane.  This is the only zero of $S$ in $\mathcal{H}$ and corresponds to the matrix model having vanishing cut length.  Below the line of degeneration $S$ has an infinite collection of zeros which appear in Figure \ref{fig:ModuliSpace2} as ``blackened" dots.  From a purely analytical perspective, all zeros and extrema of $S(\tau)$ are on the same footing, whether they are above or below the line of degeneration, which is an artifact of the matrix model.  We therefore see that $S(\tau)$ seems to exhibit a self-similarity or fractal-like structure\footnote{A possibly related remark was made in \cite{russo_massive_2013}, albeit in the context of $\mathcal{N}=2^{*}$ theory.}.  Recall that the $\mathcal{N}=1^{*}$ matrix model arose by considering a fluctuation about the \emph{confining vacuum}.  It is tempting to speculate that the structure below the line of degeneration might correspond to the other massive vacua of $\mathcal{N}=1^{*}$.  Loosely speaking, a result of \cite{dorey_massive_2002} is that solutions in the various $\mathcal{N}=1^{*}$ vacua are related by $\text{SL}(2, \mathbb{Z})$ transformations, so perhaps the self-similarity of $S$ we are observing is related to this fact.  
\end{itemize}

\noindent The above observations can be summarized in the following relation between analytical data encoded into the 't Hooft coupling $S(\tau)$ and the geometrical configuration of the eigenvalue plane:

\vskip3ex
\noindent \textbf{The unique zero of $\bm{S(\tau)}$ in $\bm{\mathcal{H}}$ corresponds to the unique point in $\bm{\mathcal{H}}$ where the matrix model has vanishing cut length.  Moreover, the points labeled $\bm{n=1,2,\ldots}$ in Figure \ref{fig:modspplot} where the mirror cuts of the matrix model are approaching degeneration with cut length $\bm{n|M|}$ correspond to the extrema $\bm{\tau_{c}^{(n)}}$ of $\bm{S(\tau)}$ for all $\bm{M \in \mathbb{C}}$.}

\section{A Relationship between the Gauge Theory and the Matrix Model}

In addition to $S$, the other important quantity arising from the matrix model computation is \cite{dijkgraaf_perturbative_2002,dorey_exact_2002-1},

\begin{equation}
\frac{\partial \mathcal{F}_{0}}{\partial S} = i \int_{B} G(x) dx,
\end{equation}
where $\mathcal{F}_{0}$ is the genus-zero part of the free energy expansion, and where $B$ is the contour on the eigenvalue plane which runs directly from $C^{-}$ to $C^{+}$.  By a computation similar to the one in the previous section, one can compute $\partial \mathcal{F}_{0}/\partial S$ and show that it is related to $S$ through,

\begin{equation}
\frac{\partial \mathcal{F}_{0}}{\partial S} = 2 \pi i \tau S -\frac{1}{12} E_{2}(\tau).
\end{equation}
The effective superpotential in the confining vacuum is then given as,

\begin{equation}
W_{\textnormal{eff}} = N \frac{\partial \mathcal{F}_{0}}{\partial S} - 2 \pi i \tau_{0} S,
\end{equation}
where $\tau_{0}$ is the bare gauge theory coupling.  One must then minimize the superpotential with respect to $\tau$, which leads to a remarkable relationship between the gauge theory coupling, and the modular parameter of the elliptic curve,

\begin{equation}
\tau = \frac{(\tau_{0} + k)}{N},
\end{equation}
where $k=0, \ldots, N-1$ labels the $N$ distinct confining vacua.  Recalling the form of the bare gauge theory coupling (\ref{eqn:gaugethcoup}), we see

\begin{equation}
\tau = \frac{k}{N} + \frac{\theta}{2 \pi N} + \frac{4 \pi i}{g_{\textnormal{YM}}^{2}N}.
\end{equation}
If we take the 't Hooft limit (in the gauge theory!) then the term involving $\theta$ drops out, but since $k$ can scale with $N$, the term $k/N$ remains.  Since $k$ is an integer between 0 and $N-1$, if $N \to \infty$, then,

\begin{equation}
0 \leq \frac{k}{N} < 1.
\end{equation}
Moreover, since $\lambda = g_{\textnormal{YM}}^{2}N$ is the gauge theory 't Hooft coupling, we arrive at the following expression for the modular parameter $\tau$ in terms of gauge theory quantities,

\begin{equation} \label{eqn:modspacecoord}
\tau = \frac{k}{N} + \frac{4 \pi i}{\lambda},
\end{equation}
where in this expression, $k/N$ is to be interpreted as a real number greater than or equal to zero, and strictly less than one.  This is consistent with the choice $0 \leq \textnormal{Re}(\tau) < 1$ in Figure \ref{fig:ModuliSpace2}.  Equation (\ref{eqn:modspacecoord}) seems to give a physical interpretation of the coordinate $\tau$ on the moduli space $\mathcal{H}$.  It expresses $\tau$ in terms of the parameters of a gauge theory in the 't Hooft limit.  The real part of $\tau$ corresponds to a choice of scaling of $k$ with $N$, while the imaginary part of $\tau$ is related to the gauge theory 't Hooft coupling $\lambda$.  $S(\tau)$ is then a holomorphic function on this space, and what I now want to show is that at weak coupling (small $\lambda$), $S(\tau)$ exactly determines the configuration of the matrix model.

\section{Weak Coupling Expansion of $S(\tau)$}

Recall that the periodicity of the Eisenstein series implies that $S(\tau)$ is also invariant under $\tau \to \tau + 1$.  Of course, it follows that $S$ can be Fourier expanded in $\tau$.  It's conventional to define the complex parameter $q = e^{2 \pi i \tau}$.  Using (\ref{eqn:modspacecoord}), we see that $q$ takes the form,

\begin{equation}
q = e^{-\frac{8 \pi^{2}}{\lambda}}e^{2 \pi i k/N}.
\end{equation}
Thus, it is clear that small $|q|$ is the region of small 't Hooft coupling $\lambda$.  This is the region of \emph{large} imaginary part of $\tau$.  The Fourier expansion takes the form

\begin{equation} \label{eqn:qexp}
S(\tau) = \sum_{n=0}^{\infty} c_{n}q^{n}.
\end{equation}
To lowest order in $q$

\begin{equation}
S(\tau) = -2q + \mathcal{O}(q^{2}),
\end{equation}
and substituting in the form of $q$ above (\ref{eqn:qexp}), we see

\begin{equation}
S(\tau) \approx 2 e^{-\frac{8 \pi^{2}}{\lambda}} e^{2 \pi i(k/N + 1/2)}.
\end{equation}
This holomorphic function on $\mathcal{H}$ determines precisely the configuration of the matrix model at weak coupling.  More specifically, the configuration of the matrix model is given simply by providing the angle of the cut, and the length of the cut.  The function, which to my knowledge has not appeared in the literature,

\begin{equation} \label{eqn:MMConfig}
\boxed{Q(\tau) = -\sqrt{8 M^{2} S(\tau)}}
\end{equation}
has a Fourier expansion which I hope to show recovers the matrix model branch cut configuration at weak coupling.  The lowest order term in the expansion is,

\begin{equation}
 Q(\tau) \approx -\bigg(4 |M| e^{-\frac{4 \pi^{2}}{\lambda}}\bigg) e^{i \pi(k/N + 1/2) + i \textnormal{Arg}(M)}.
\end{equation}
Restricting to the Hermitian slice, we can choose $\frac{k}{N}=\frac{1}{2}$ and $M \in \mathbb{R}$.  In such a case, the above expansion simplifies to

\begin{equation} \label{eqn:weakcoupres}
Q \approx 4M e^{-4 \pi^{2}/\lambda}, \,\,\,\,\,\,\,\,\,\,\,\,\,\,\,\,\,\, \big(k/N=1/2, \,\, M \in \mathbb{R}\big).
\end{equation}
We've seen essentially the same equation before (\ref{eqn:weakcutleng}) in the context of weak coupling cut lengths in the $\mathcal{N}=2^{*}$ matrix model.  Recalling that equation (\ref{eqn:weakcutleng}) refers to the \emph{half}-cut length, the lowest order term in the Fourier expansion of $Q(\tau)$ restricted to the Hermitian slice appears to encode the full cut length $2 \mu$ of $\mathcal{N}=2^{*}$ at weak coupling.  Similarly, choosing the anti-Hermitian slice defined by $\frac{k}{N}=0$ along with $M \in \mathbb{R}$, the resulting expansion has the same modulus as (\ref{eqn:weakcoupres}) but with an additional factor of $-i$.  It is therefore natural to speculate that perhaps at weak coupling, the modulus of $Q(\tau)$ encodes the matrix model cut length and its argument encodes the cut angle.  I now want to provide graphical evidence that this is indeed the case.

\begin{figure}
\begin{center}
\includegraphics[width=10cm]{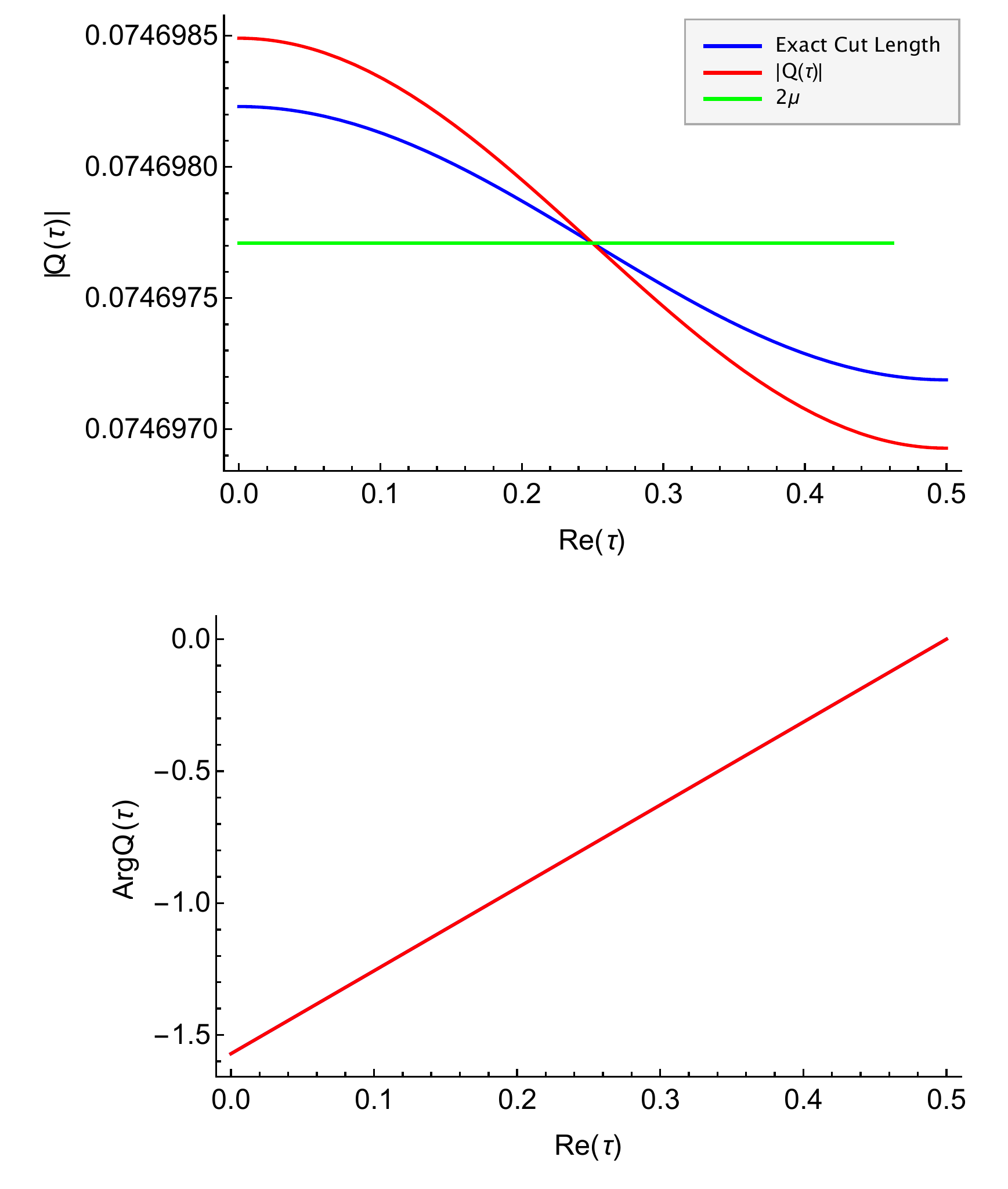}
\caption{$\bm{(\lambda = 2 \pi, M=10)}$ Along this horizontal slice, the upper plot shows the matrix model cut length (blue), the modulus of $Q$ (red), and $2\mu$ (green).  It appears that $|Q|$ encodes very closely the variation of the matrix model cut length across the slice.  If we were to make $\lambda$ even smaller, the blue and red curves would converge to $2\mu$.  This is the statement that at incredibly small coupling, the horizontal slices across $\mathcal{H}$ correspond to slices of constant cut length.  The second plot actually shows both the matrix model cut angle and the argument of $Q$, but the match is so precise that the two are indistinguishable.  It is reasonable to conclude from these two plots that at extremely small $\lambda$, the horizontal slices through the moduli space correspond to constant cut length and linearly increasing cut angle. \label{fig:WeakCoupData1}}
\end{center}
\end{figure}

\begin{figure}
\begin{center}
\includegraphics[width=10cm]{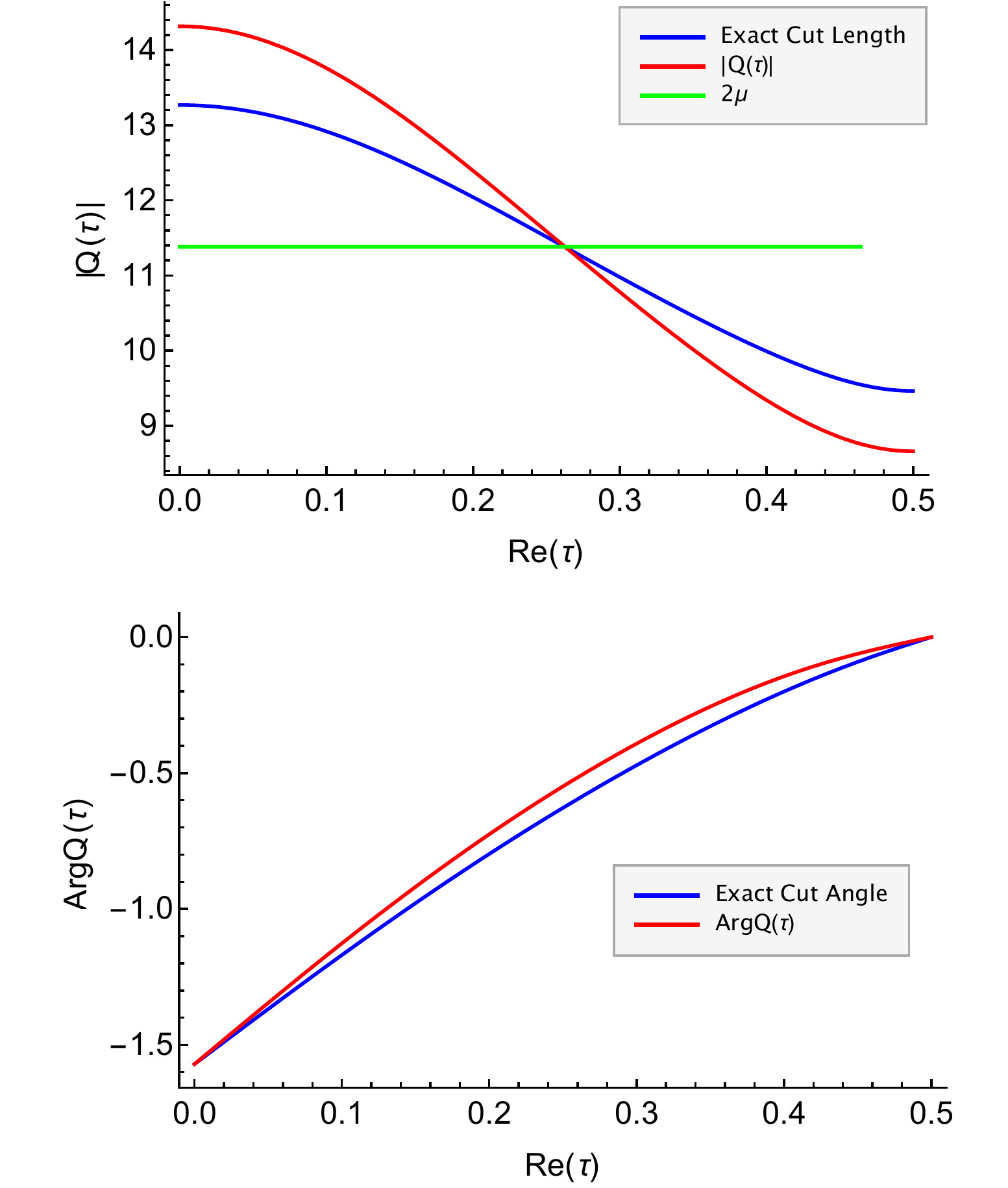}
\caption{$\bm{(\lambda = 31.416, M=10)}$ Also along a horizontal slice, but in this case at larger coupling $\lambda = 31.416$.  Notice now that $|Q(\tau)|$ deviates more appreciably from the actual cut length.  In addition, $\textnormal{Arg}Q(\tau)$ no longer perfectly fits the cut angle.  Though the two agree at the endpoints of the slice, the actual cut angle varies more linearly than $\textnormal{Arg}Q(\tau)$ predicts.  This is the first clue that although $S(\tau)$ is deeply connected to the matrix model, it may lose contact with the exact configuration at strong coupling.  \label{fig:WeakCoupData2}}
\end{center}
\end{figure}

\begin{figure}
\begin{center}
\includegraphics[width=11cm]{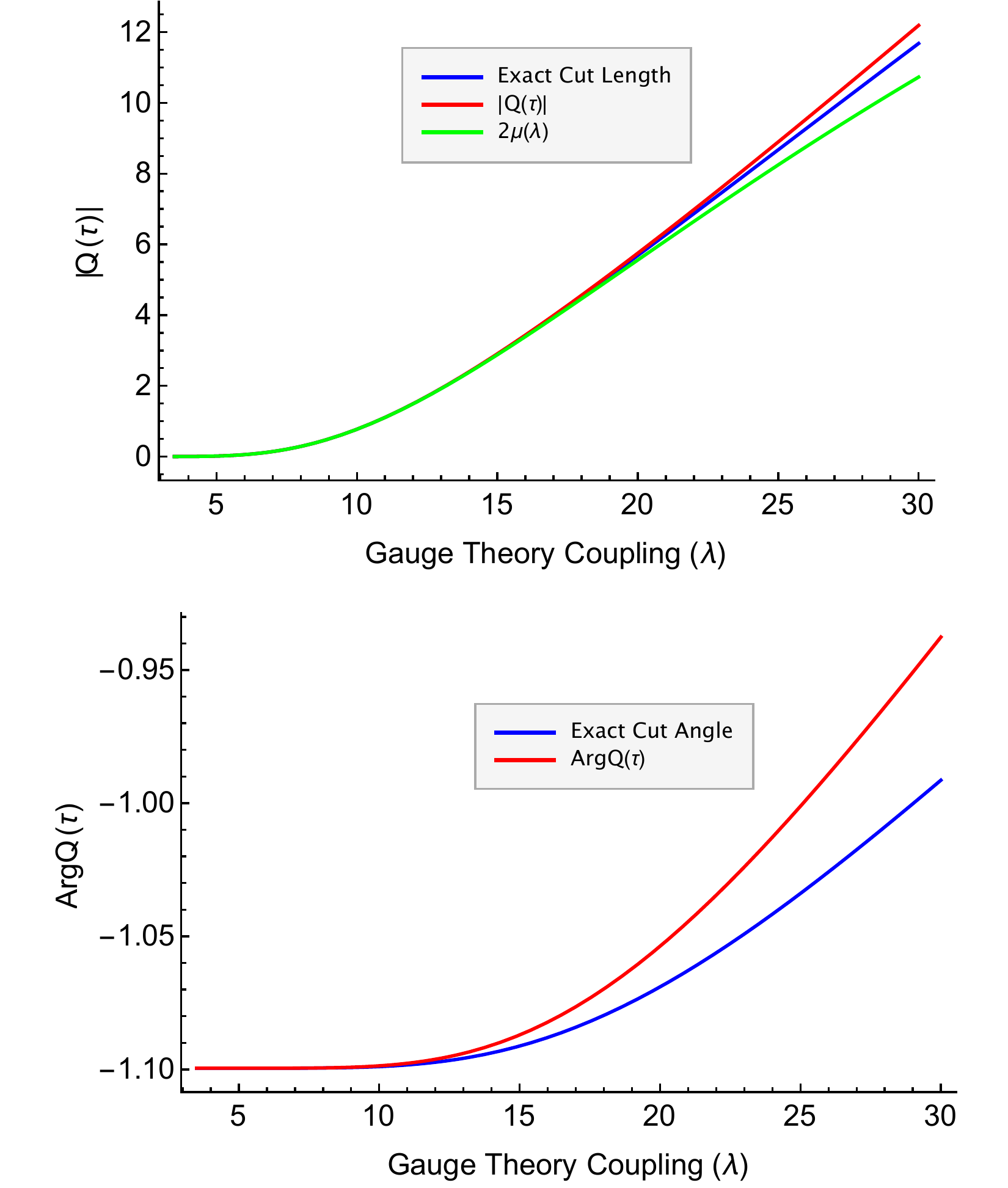}
\caption{$\bm{(\textnormal{Re}(\tau)=0.15, M=10)}$ One can imagine instead fixing $\textnormal{Re}(\tau)$ and considering a range of $\lambda$.  As we can see, at weak coupling $2 \mu(\lambda) = 4M e^{-4 \pi^{2}/\lambda}$, $|Q(\tau)|$, and the actual cut length agree fantastically.  As the coupling increases, the three begin deviating.  Moreover, the cut angle agrees at weak coupling with $\text{Arg}Q(\tau)$ and both are approximately constant, but around $\lambda = 15$, $\textnormal{Arg}Q(\tau)$ begins growing much faster than the cut angle.  Thus, at weak coupling on a vertical slice, the cut length obeys the exponential relationship $2 \mu(\lambda) = 4Me^{-4 \pi^{2}/ \lambda}$ while the cut angle remains fixed.  Conversely, recall that on a horizontal slice at weak coupling the cut length was constant in the strict $\lambda \to 0$ limit, while the cut angle varied linearly. \label{fig:tHooftPlot1}}
\end{center}
\end{figure}

\begin{figure}
\begin{center}
\includegraphics[width=11cm]{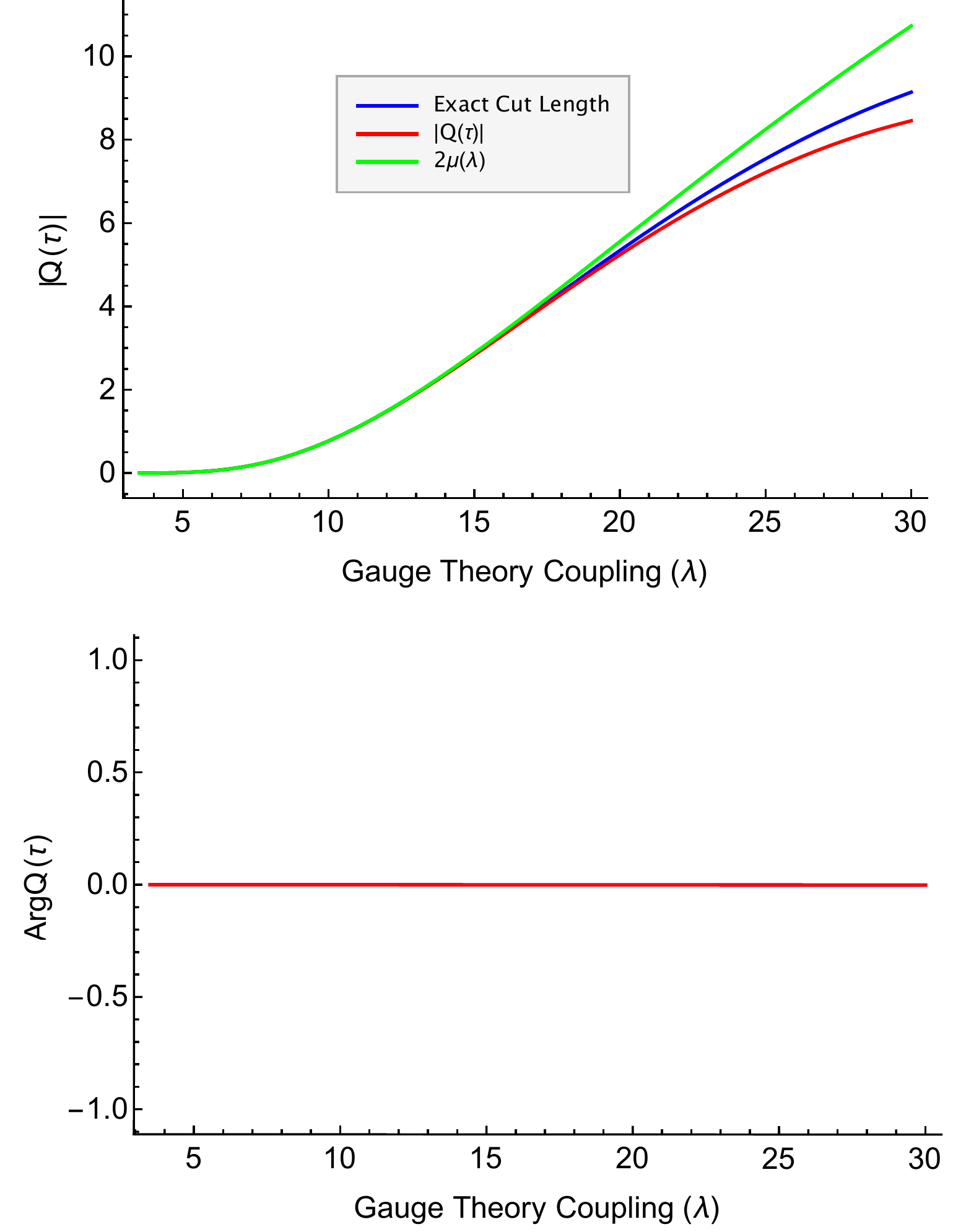}
\caption{$\bm{(\textnormal{Re}(\tau)=0.5, M=10)}$ As in the previous figure, we fix $\textnormal{Re}(\tau)$ and consider a range of $\lambda$.  This corresponds to the Hermitian slice above the critical point.  Notice that the upper plot is very similar to that in Figure \ref{fig:tHooftPlot1}.  However, along this slice both the exact cut angle and $\text{Arg}Q(\tau)$ are identically zero until the mirror cuts collide at the origin and the model degenerates. \label{fig:tHooftPlot2}}
\end{center}
\end{figure}

Within $\mathcal{H}$, there are two simple types of slices we can look at.  We can fix $\lambda$, and run horizontally across the range $0 \leq \textnormal{Re}(\tau) < 1/2$ (it suffices to look at half of $\mathcal{H}$ by symmetry).  In Figures \ref{fig:WeakCoupData1} and \ref{fig:WeakCoupData2} we analyze slices of this form for two different couplings.  Conversely, we can fix $\textnormal{Re}(\tau)$ and consider a range of 't Hooft couplings.  Such slices are analyzed in Figures \ref{fig:tHooftPlot1} and \ref{fig:tHooftPlot2}.

\section{Surprising Connections to $\mathcal{N}=2^{*}$ Theory}

The computation I present at the beginning of this chapter results in a quantity $S(\tau)$, and hence $Q(\tau)$, which is intrinsically an object in $\mathcal{N}=1^{*}$ theory.  Nevertheless, I argue here that these quantities actually encode non-trivial information about $\mathcal{N}=2^{*}$ theory along with its sequence of critical points.  I have already stated explicitly the following conclusion which I repeat here for completeness:

\vskip3ex
\noindent \textbf{For all $\bm{M \in \mathbb{R}}$, the weak coupling expansion of $\bm{Q(\tau)}$ restricted to the Hermitian slice through $\bm{\mathcal{H}}$ exactly recovers the well-known weak coupling cut length in $\bm{\mathcal{N}=2^{*}}$ theory.}
\vskip3ex

\noindent This is an example of non-trivial $\mathcal{N}=2^{*}$ data encoded into $\mathcal{N}=1^{*}$ quantities.  In addition, there was actually a hint earlier in the chapter of a connection to $\mathcal{N}=2^{*}$.  Recall that the extrema $\tau_{c}^{(n)}$ of $S(\tau)$ encode the locations in $\mathbb{H}$ where the two mirror cuts are approaching degeneration, each with length $n |M|$ for all $M \in \mathbb{C}$.  In my brief review of the $\mathcal{N}=2^{*}$ matrix model, I state the result of \cite{russo_massive_2013} for the \emph{half}-cut lengths at the critical points, which I recall here

\begin{equation}
\mu(\lambda_{c}^{(n)}) = \frac{nM}{2}.
\end{equation}
In this context, $M \in \mathbb{R}$ and $\lambda_{c}^{(n)}$ is the gauge theory 't Hooft coupling at the $n$-th critical point.  Multiplying by 2 to account for the full cut length, we see that indeed the critical cut lengths in $\mathcal{N}=2^{*}$ theory appear to arise in my model at the extrema of $S(\tau)$.  

There are however a number of issues.  First, in \cite{russo_massive_2013} where Russo and Zarembo first apply the modular solution to the $\mathcal{N}=2^{*}$ matrix model, they note that upon encountering the first critical point ($n=1$) the modular solution degenerates altogether.  Abandoning the modular solution, Russo and Zarembo numerically push beyond the frontier of the $n=1$ critical point, allowing the matrix model cut to expand freely on the real axis, without any reference to the mirror cuts of the modular solution.  Doing so, they numerically find an infinite sequence of quantum critical points with critical 't Hooft couplings $\lambda_{c}^{(n)}$ accumulating at infinite coupling.  

One of my original hopes was that the holomorphic nature of my model might be able to detect these quantum critical points \emph{analytically}.  Perhaps the strong coupling region only appeared inaccessible to one who was constraining themselves to a Hermitian slice.  Approaching the extrema $\tau_{c}^{(n)}$ of $S(\tau)$ from within $\mathcal{H}$, indeed the branch cut is converging to exactly the critical configuration of Russo and Zarembo.  Thanks to the above construction of Dijkgraaf and Vafa, we can coordinatize the moduli space using gauge theory parameters $\frac{k}{N}$ and $\lambda$ (\ref{eqn:modspacecoord}), which relates the imaginary part of $\tau$ to $\lambda$.  I want to compare $\lambda_{c}^{(n)}$ with $\tilde{\lambda}_{c}^{(n)}$, where $\lambda_{c}^{(n)}$ are the $\mathcal{N}=2^{*}$ critical couplings computed by Russo and Zarembo \cite{russo_massive_2013,russo_evidence_2013,hollowood_partition_2015}, and $\tilde{\lambda}_{c}^{(n)}$ are my ``predicted" couplings defined by

\begin{equation}
\textnormal{Im}(\tau_{c}^{(n)}) = \frac{4 \pi}{\tilde{\lambda}_{c}^{(n)}},
\end{equation}
where $\tau_{c}^{(n)}$ are the extrema of $S(\tau)$.  My parameter values are reported in Table \ref{Table:CritPointsTab11} below.

\begin{table}
\centering
\begin{tabular} { |c|c|c| }
\hline
               & $\bf{\tau_{c}^{(n)}}$ & $\bf{\tilde{\lambda}_{c}^{(n)}}$ \\ \hline
$\,\,\,\,\,\,\,\,\,\,\bf{n=1}\,\,\,\,\,\,\,\,\,\,$ & $\,\,\,\,\,\,\,\,\,\,0.5+0.35473\,i\,\,\,\,\,\,\,\,\,\,$              & \,\,\,\,\,\,\,\,\,\,$35.4252\ldots$\,\,\,\,\,\,\,\,\,\,                           \\ \hline
$\,\,\,\,\,\,\,\,\,\,\bf{n=2}\,\,\,\,\,\,\,\,\,\,$ & $\,\,\,\,\,\,\,\,\,\,0.345+0.16075\,i\,\,\,\,\,\,\,\,\,\,$              & \,\,\,\,\,\,\,\,\,\,$78.1734\ldots$\,\,\,\,\,\,\,\,\,\,                           \\ \hline
$\,\,\,\,\,\,\,\,\,\,\bf{n=3}\,\,\,\,\,\,\,\,\,\,$ & $\,\,\,\,\,\,\,\,\,\,0.2605+0.0928\,i\,\,\,\,\,\,\,\,\,\,$              & \,\,\,\,\,\,\,\,\,\,$135.413\ldots$\,\,\,\,\,\,\,\,\,\,                           \\ \hline
$\,\,\,\,\,\,\,\,\,\,\bf{n=4}\,\,\,\,\,\,\,\,\,\,$ & $\,\,\,\,\,\,\,\,\,\,0.209+0.0602\,i\,\,\,\,\,\,\,\,\,\,$              &\,\,\,\,\,\,\,\,\,\, $208.744\ldots$\,\,\,\,\,\,\,\,\,\,                           \\ \hline
$\,\,\,\,\,\,\,\,\,\,\bf{n=5}\,\,\,\,\,\,\,\,\,\,$ & $\,\,\,\,\,\,\,\,\,\,0.1735+0.0429 \, i\,\,\,\,\,\,\,\,\,\,$              &\,\,\,\,\,\,\,\,\,\, $292.922\ldots$\,\,\,\,\,\,\,\,\,\,                           \\ \hline
\end{tabular}
\caption{The first five critical points of $S(\tau)$ and the corresponding gauge theory couplings.  My value of $\tilde{\lambda}_{c}^{(1)}$ agrees perfectly with \cite{russo_massive_2013,russo_evidence_2013,hollowood_partition_2015}, however the higher 't Hooft couplings disagree with the numerical predictions in \cite{russo_massive_2013,russo_evidence_2013}. \label{Table:CritPointsTab11}}
\end{table}

\noindent The agreement is exact for $n=1$!  One can check that $\tilde{\lambda}_{c}^{(1)} = \lambda_{c}^{(1)}$ which means that the $n=1$ extrema of $S(\tau)$ encodes the $n=1$  critical coupling of $\mathcal{N}=2^{*}$ theory.  Unfortunately, the couplings \emph{do not} match for $n>1$.  

A relevant result of \cite{hollowood_partition_2015} is that the $\mathcal{N}=2^{*}$ theory is ill-defined off the slice $\frac{k}{N}=\frac{1}{2}$ and that the quantum critical points indeed lie on this slice, with increasingly small imaginary part.  Such points lie in the region of degeneration of my model, so it appears that maybe the modular solution to $\mathcal{N}=2^{*}$ genuinely cannot detect the critical points for $n>1$.  However, I offer a highly speculative potential resolution (which I am unable to check): \emph{what if the quantum critical points on the Hermitian slice can be taken to the extrema of $S(\tau)$ by a modular transformation?}  Perhaps there is an anomalous term in the relationship between $\tau$ and $\lambda$ arising when performing a modular transformation, which would account for the above discrepancy in the couplings.  The extrema of $S(\tau)$ are reminiscent enough of the quantum critical points to deserve an explanation.  If one were to abandon the modular solution of $\mathcal{N}=2^{*}$ altogether, then the single branch cut centered at the origin could freely elongate past the first critical point along the real axis, sequentially hitting cut lengths which are integer multiples of $M$.  In my model, upon collision with the first critical point, the rest of the $\frac{k}{N}=\frac{1}{2}$ slice is off-limits.  Instead, one could travel along (rather, arbitrarily closely to) the line of degeneration within the bulk of $\mathcal{H}$.  Here, the cut lengths approaching integral multiples of $M$ would be detected by the sequence of extrema of $S(\tau)$ along this line.  

All speculations aside, I can at least conclude the following from the above discussion:

\vskip2ex
\noindent \textbf{Under the change of variables $\text{Im}\bm{(\tau) = 4 \pi/ \lambda}$, the $\bm{n=1}$ extrema of $\bm{S(\tau)}$ exactly encodes the first critical point of $\bm{\mathcal{N}=2^{*}}$ theory.}
\vskip2ex

\noindent I want to emphasize that all of the connections to $\mathcal{N}=2^{*}$ I have presented in this section, both exact and speculative, would be \emph{invisible} if one were to restrict the $\mathcal{N}=1^{*}$ theory to the anti-Hermitian slice $\frac{k}{N}=0$.  It is the true holomorphic nature of the model developed in the last two chapters which reveal that $\mathcal{N}=1^{*}$ encodes exact weak coupling cut lengths in $\mathcal{N}=2^{*}$, as well as at least, the critical 't Hooft coupling $\lambda_{c}^{(1)}$.

\chapter{Eigenvalue Densities}

As a purely geometrical problem, we've seen that given any $\tau \in \mathcal{H}$ and $M \in \mathbb{C}$ we can determine the configuration of the branch cuts in the eigenvalue plane as well as the non-trivial cycle on the elliptic curve mapping into the cut.  This construction contained no physics.  I now want to compute eigenvalue densities in the $\mathcal{N}=1^{*}$ and $\mathcal{N}=2^{*}$ matrix models.  Eigenvalue densities clearly depend on the physical theory one is working in, and therefore naturally should depend on the choice of generalized resolvent.  We've encountered the generalized resolvent $G(x(z))$ of $\mathcal{N}=1^{*}$ before (\ref{eqn:N1starresolve}) as computed in \cite{dorey_exact_2002-1}.  Recall,

\begin{equation}\label{eqn:secondGx}
G(x(z)) = \frac{M^{2}}{4} \bigg[\wp(z) - \frac{2}{3} E_{2}(\tau)\bigg].
\end{equation}
The generalized resolvent of the $\mathcal{N}=2^{*}$ theory was provided in \cite{hollowood_partition_2015},

\begin{equation} \label{eqn:N2starresolve}
\widetilde{G}(x(z)) = \frac{4}{M} \frac{1}{\wp(z) + \tfrac{1}{3}E_{2}(\tau)}.
\end{equation}
In this chapter I give a prescription for recovering eigenvalue densities, which is not explicit in the literature as far as I am aware.  Given a choice of generalized resolvent, the eigenvalue densities are encoded into the restriction of this elliptic function to the cycle on the elliptic curve mapping into the cut.  At weak coupling, my method recovers the inverse square-root density of $\mathcal{N}=2^{*}$ as well as Wigner's semi-circular density in $\mathcal{N}=1^{*}$.  There are numerical and asymptotic indications in the literature that the $\mathcal{N}=1^{*}$ matrix model in the limit of large cut length supported in the real axis should admit a parabolic density.  Of course, this region lies on the line of degeneration in my model.  \emph{Making use of the holomorphic nature of the $\mathcal{N}=1^{*}$ model}, I attempt to find evidence of the parabolic density in the region of $\mathcal{H}$ where the two mirror cuts are elongated and approaching an overlapping configuration on the real axis.  Indeed, one of the original motivations of this project was to see if the elliptic curve and generalized resolvent encoded the parabolic density, and I present partial evidence in favor of this.

\section{The General Method}

As we have seen, for each $\tau \in \mathcal{H}$ there exists a map $x(z)$ which determines the configuration of the eigenvalue plane.  In addition, there must exist a distinguished $A$-cycle denoted $\mathcal{C}^{+}$ which maps into the cut $C^{+}$ on the eigenvalue plane with midpoint $\tfrac{M}{2}$.  In other words, the restriction of the map to $\mathcal{C}^{+}$ takes values in the cut itself.  Subtracting $\tfrac{M}{2}$ we map to the \emph{real} cut whose midpoint is the origin (recall the two cuts are the ``mirrors" of the single real cut).  We choose to parameterize $\mathcal{C}^{+}$ with a real parameter $0 \leq s \leq 2$, such that the points in the real cut are $x(s)-\tfrac{M}{2}$, for all $s$.

\begin{figure}
\begin{center}
\includegraphics[width=8cm]{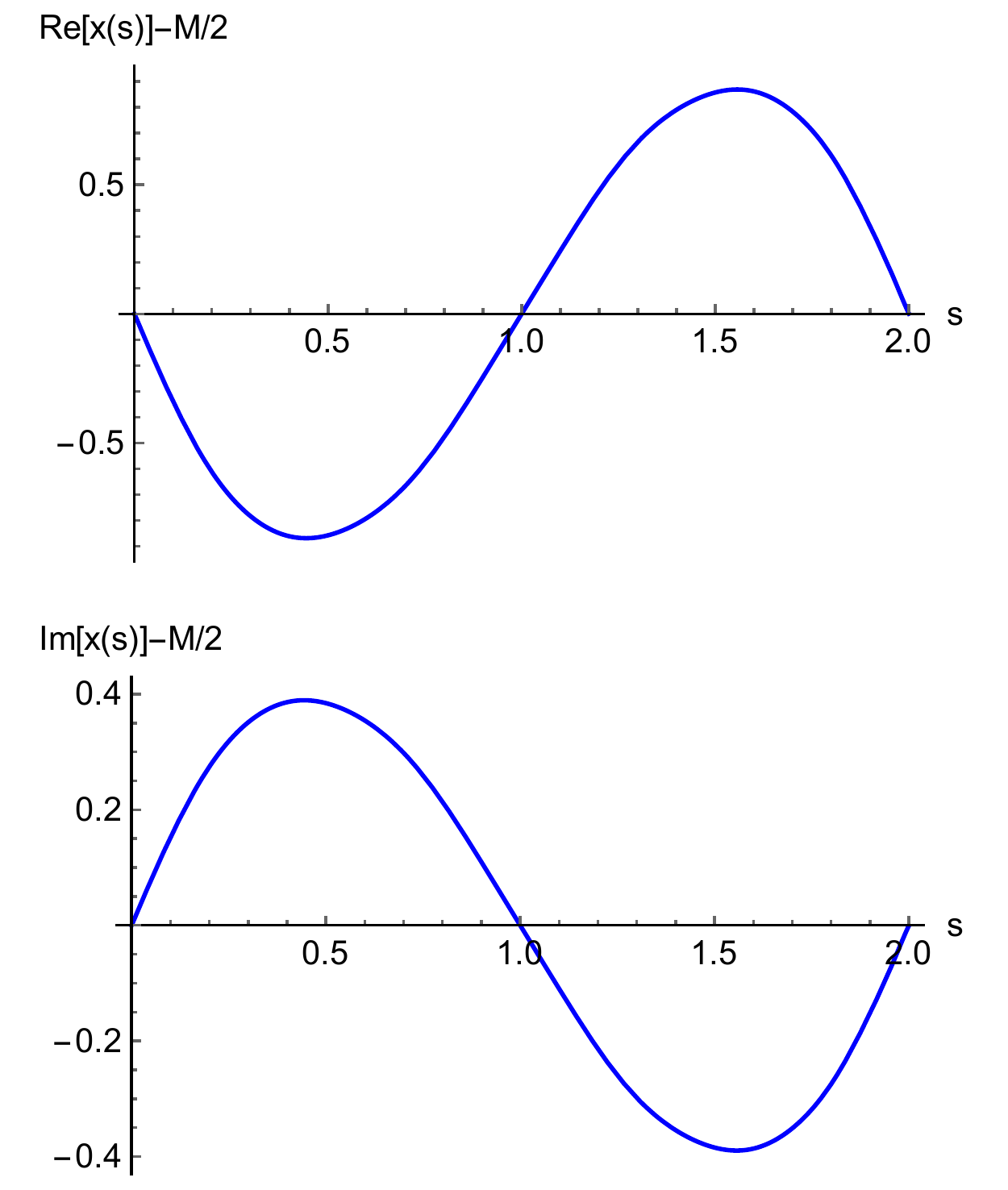}
\caption{For $\tau = \frac{1}{4}+ \frac{1}{4} i$, the map from $\mathcal{C}^{+}$ into the real and imaginary parts of the cut $C^{+}$. \label{fig:embres}}
\end{center}
\end{figure}
In Figure \ref{fig:embres} above, for $s=0$ we begin at the origin (i.e. the midpoint of the cut), then proceed to maximally negative real part, maximally positive imaginary part, turn around, traverse the cut in the other direction, and finally return to the origin.  The $s$ values at which the maxima and minima occur parameterize the distinguished points in $\mathcal{C}^{+}$ mapping to the branch points.  Indeed, both the real and imaginary plots have these extrema occurring at the same parameter values, as they must.  Finally, notice that the range of the real part is slightly larger than the range of the imaginary part.  Looking back at Figure \ref{fig:cutgeo1} shows this is because the cut is extended more along the real axis than the imaginary axis.  Recall the closed form expression for the eigenvalue density in terms of the discontinuity of the ordinary resolvent across the branch cut,

\begin{equation} \label{eqn:eigdensfinal}
\rho(x) = -\frac{1}{2 \pi i }\bigg(\omega(x + i \epsilon)- \omega(x - i \epsilon)\bigg).
\end{equation}
Though the generalized resolvents may appear in a different form than the ordinary resolvents, when it comes to discontinuities across cuts, they encode the same data.  As such, the plan is to exploit the relationship of the generalized resolvent to the ordinary resolvent, and then in turn, the relationship of the ordinary resolvent to the eigenvalue density.  This will yield an exact method for extracting eigenvalue densities.

Since the cut is a straight line segment, we can parameterize its points just as well by their real parts, imaginary parts, or by a length parameter along the cut.  In this context, it is most natural to choose a length parameter.  Let $\eta$ be the real part of a point in the cut.  We define a length parameter $\xi$ by $\eta = \xi \cos \theta$, where $\theta$ is the angle of the cut with respect to the real axis.  Clearly, the parameter $\xi$ spans the full length of the cut, instead of merely the spanning the real or imaginary parts.  The cycle $\mathcal{C}^{+}$ double covers the cut $C^{+}$.  This makes sense if we recall that the image of $A$-cycles under $x$ encircle the branch cut.  The image $x(\mathcal{C}^{+})$ does so ``infinitely tightly", which we think of as a double cover.  Therefore:
\vskip3ex
\noindent \textbf{for all $\bm{\xi}$ in the branch cut, there exists two points $\bm{s_{1}^{\xi}}$ and $\bm{s_{2}^{\xi}}$ in $\bm{\mathcal{C}^{+}}$ mapping by $\bm{x}$ to the same point in the cut with parameter $\bm{\xi}$.  Hence, the discontinuity of the resolvent across the cut can be reformulated as the difference in the two values of the generalized resolvent evaluated at these two points on the elliptic curve.  This provides the value of the eigenvalue density at this single point of the cut.}
\vskip3ex

\noindent Sweeping through each point in the cut, I can apply this idea to compute the exact eigenvalue density as a function of $\xi$ along the length of the cut for all $\tau \in \mathcal{H}$.  Exact expressions for the eigenvalue densities are

\begin{equation}
\boxed{\rho(\xi)=
\begin{cases}
\frac{1}{2 \pi i M^{3} S} \bigg(G\big(x(s_{1}^{\xi})\big) - G\big(x(s_{2}^{\xi})\big)\bigg), & \,\,\,\,\,\,\,\, \mathcal{N}=1^{*}\\
-\frac{1}{2 \pi i} \bigg(\widetilde{G}\big(x(s_{1}^{\xi})\big) - \widetilde{G}\big(x(s_{2}^{\xi})\big)\bigg), &\,\,\,\,\,\,\,\, \mathcal{N}=2^{*}
\end{cases}}
\end{equation}
This is rather meager as far as self-contained formulae go, given that for each $\xi$ one must do a computation to find $s_{1}^{\xi}$ and $s_{2}^{\xi}$.  But it nonetheless shows explicitly that the eigenvalue densities are encoded into the elliptic curve and the generalized resolvent, the exact form of which, is known.


Note that $s_{1}^{\xi} = s_{2}^{\xi}$ precisely when we are at an endpoint of the cut.  By construction, we've seen the transcendental constraint (\ref{eqn:transeq}) which must hold at an endpoint, 

\begin{equation} 
\wp(s_{1}^{\xi}) = \wp(s_{2}^{\xi}) = -\frac{1}{3}E_{2}(\tau).  
\end{equation}
From (\ref{eqn:secondGx}), we see that $G(x(s_{1})) = G(x(s_{2}))$ and thus, the difference will vanish.  This is the statement that the eigenvalue densities in the $\mathcal{N}=1^{*}$ theory vanish at the endpoints.  On the other hand, the denominator of (\ref{eqn:N2starresolve}) implies that both $\widetilde{G}(x(s_{1}))$ and $\widetilde{G}(x(s_{2}))$ diverge.  So we expect eigenvalue densities in the $\mathcal{N}=2^{*}$ theory to diverge at the endpoints.  This is well-known from \cite{russo_massive_2013}, and in fact is what motivated the construction in \cite{hollowood_partition_2015}.

\subsection*{Reality and Normalization of Eigenvalue Densities?}

A final remark is in order before presenting results.  As noted in \cite{lazaroiu_holomorphic_2003-1}, eigenvalue densities in holomorphic matrix models are generally complex.  Indeed, this is true of my construction.  For general $\tau \in \mathcal{H}$, the resulting eigenvalue density will be complex valued except along distinguished ``slices" through the moduli space.  For example, the eigenvalue density is real along the entire anti-Hermitian slice $\textnormal{Re}(\tau)=0$.  In addition, along the Hermitian slice $\textnormal{Re}(\tau)=1/2$ (above the critical point) the density is also real valued.

Given that the eigenvalue densities are complex, what we mean by normalization must be adjusted.  I have checked in a large number of cases that it is precisely the \emph{absolute value} of the eigenvalue densities $| \rho(\xi)|$ which is normalized over the full length of the cut.  This is true of both $\mathcal{N}=1^{*}$ and $\mathcal{N}=2^{*}$.  Therefore, when the imaginary part vanishes, the eigenvalue densities will satisfy both reality and normalization conditions.  In my recovery of the semi-circle and the inverse square-root, the eigenvalue densities will be real and normalized perfectly.  However, in the bulk of the moduli space $\mathcal{H}$, where I have observed evidence of the parabolic density, I must take the absolute value of the complex eigenvalue density.

\section{$\mathcal{N}=2^{*}$ at Weak Coupling: Shenker's Inverse Square Root}

\begin{figure}
\begin{center}
\includegraphics[width=12cm,height=9cm]{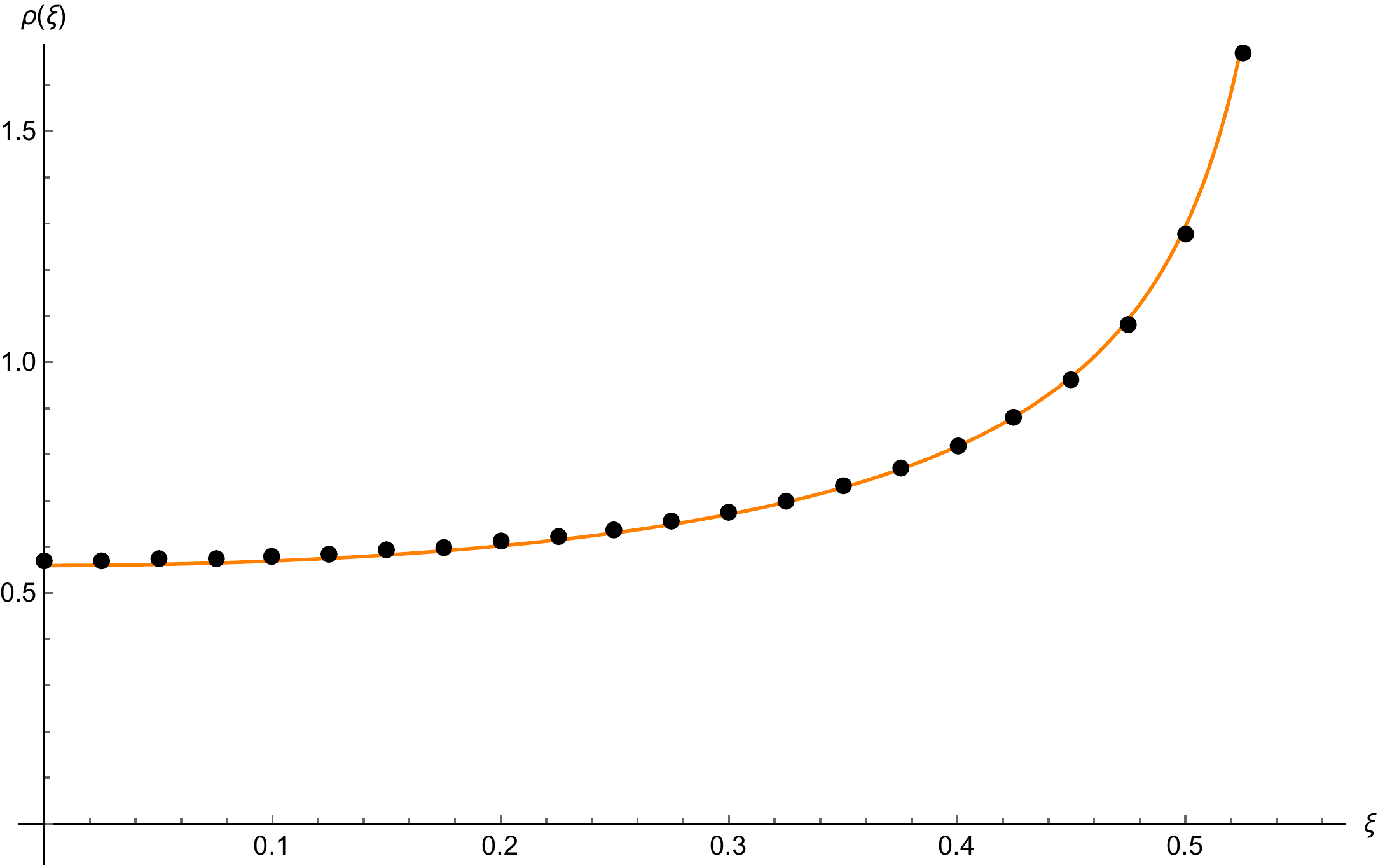}
\caption{The exact $\mathcal{N}=2^{*}$ inverse square-root eigenvalue density for $\tau = \frac{1}{2}+\frac{3}{4}i$, which corresponds to $\lambda =  \frac{16 \pi}{3} \approx 16.76$.  I plot the eigenvalue density computed using my method (orange) alongside the function (black-dotted) in (\ref{eqn:invsqqrrt}).  For the reader's convenience, this occurs at point A in Figure \ref{fig:modspplot}.  I also note that for this choice of $\tau$, the eigenvalue density is real and normalized. \label{fig:InvSqrt1}}
\end{center}
\end{figure}

We've seen that we expect the eigenvalue densities in the $\mathcal{N}=2^{*}$ to diverge at the cut endpoints.  In particular, we saw in (\ref{eqn:eigdens}) that we expect to recover Shenker's \emph{normalized} inverse square-root at weak coupling \cite{douglas_dynamics_1995}.  Choosing $\tau = \frac{1}{2} + \frac{3}{4}i$, as well as the generalized resolvent $\widetilde{G}(x(z))$ in (\ref{eqn:N2starresolve}), I plot the eigenvalue density in Figure \ref{fig:InvSqrt1}.  The normalized inverse square-root I compare to is

\begin{equation} \label{eqn:invsqqrrt}
y(\xi) = \frac{1}{\pi \sqrt{\mu^{2}-\xi^{2}}},
\end{equation}
where $\mu$ is the matrix model cut length.  Along this slice of the moduli space, the eigenvalue density is inherently real.  By symmetry, the eigenvalue density must be symmetric under inversion about the midpoint of the cut.  Thus, it suffices to plot the density over only the half-cut length $[0, \mu]$.  As we can see in Figure \ref{fig:InvSqrt1}, the eigenvalue density matches the inverse square-root remarkably well considering $\lambda \approx 16.76$ is not terribly small.  The match becomes more exact as $\lambda$ decreases.

\section{$\mathcal{N}=1^{*}$ Eigenvalue Densities}

Recall the form of the $\mathcal{N}=1^{*}$ matrix model,

\begin{equation}
\mathcal{Z}_{N}(g_{s}) = \int \mathcal{D} \Phi_{1} \mathcal{D} \Phi_{2}   \mathcal{D} \Phi_{3}  e^{-\frac{1}{g_{s}} \textnormal{Tr}\big( \Phi_{1} [ \Phi_{2}, \Phi_{3}] + \Phi_{1}^{2}+ \Phi_{2}^{2}+ \Phi_{3}^{2}\big)}.
\end{equation}
This is equation (\ref{eqn:gsMM}) written in terms of $\Phi_{i}$ instead of $\Phi$, $\Phi^{\pm}$.  By completing the square in the exponent we can integrate over $\Phi_{1}$,

\begin{equation}
\mathcal{Z}_{N}(g_{s}) = \mathcal{C} \int \mathcal{D} \Phi_{2}   \mathcal{D} \Phi_{3}  e^{-\frac{N}{S} \textnormal{Tr}\big( \Phi_{2}^{2}+ \Phi_{3}^{2} -\frac{1}{4} [\Phi_{2}, \Phi_{3}]^{2}\big)},
\end{equation}
where $\mathcal{C}$ is a complex number, and we've used $g_{s} = S/N$.  Since the 't Hooft coupling is a finite parameter, we can scale $\Phi_{2}$ and $\Phi_{3}$ by $\sqrt{S}$, resulting in,
\begin{equation} \label{eqn:FilOconn}
\mathcal{Z}_{N}(g_{s}) = \mathcal{C} \int \mathcal{D} \Phi_{2}   \mathcal{D} \Phi_{3}  e^{-N \textnormal{Tr}\big( \Phi_{2}^{2}+ \Phi_{3}^{2} -\frac{S}{4} [\Phi_{2}, \Phi_{3}]^{2}\big)}.
\end{equation}
This two-matrix model is identical to the one given in equation (2.1) in \cite{filev_multi-matrix_2013}, provided one identifies the above coupling $S/4$ with the square of theirs.  In \cite{filev_multi-matrix_2013}, Filev and O'Connor note that for strictly vanishing coupling $S=0$, we get a two-matrix model which is totally decoupled; it's merely a product of two Gaussian matrix models.  So integrating over one of them, we have simply a Gaussian one-matrix model which should produce a Wigner semi-circular eigenvalue distribution \cite{wigner_characteristic_1993}.

\subsection*{Weak Coupling: The Wigner Semi-Circle}
\begin{figure}
\begin{center}
\includegraphics[width=11.5cm]{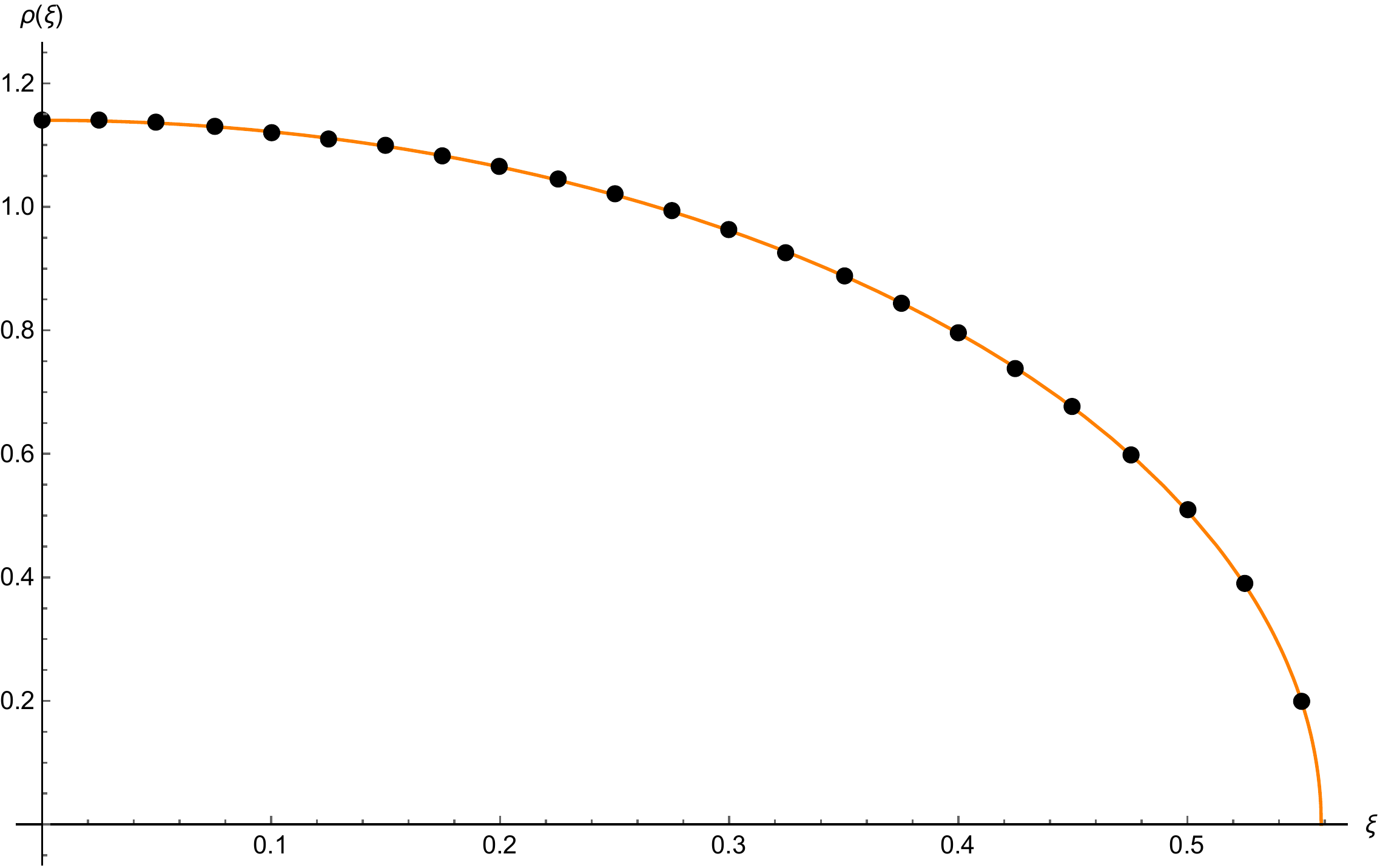}
\caption{The exact $\mathcal{N}=1^{*}$ semi-circular eigenvalue density for $\tau = \frac{1}{2}+\frac{3}{4} i$, which corresponds to $\lambda = \frac{16 \pi}{3} \approx 16.76$.  I plot the exact eigenvalue density (orange) alongside the semi-circle (black-dotted) given in (\ref{eqn:semcirc}).  This occurs at point A in Figure \ref{fig:modspplot}.  This eigenvalue density is real-valued and normalized. \label{fig:SemiCircle1}}
\end{center}
\end{figure}

Indeed, I show that even at small, but non-vanishing coupling $S$, my method recovers the Wigner semi-circle.  Recalling that we use $\xi$ as a parameter on the cut, the normalized semi-circle to which I will compare my plot is,

\begin{equation}\label{eqn:semcirc}
y(\xi) = \frac{1}{\mu \pi} \sqrt{1 - \frac{\xi^{2}}{\mu^{2}}},
\end{equation}
where $\mu$ is the matrix model cut length.  Let $\tau = \frac{1}{2} + \frac{3}{4}i$, which corresponds to a gauge theory 't Hooft coupling $\lambda = \frac{16 \pi}{3}$.  Recall that on this portion of the moduli space, the cut lives entirely in the real axis.  Notice from Figure \ref{fig:SemiCircle1} that the fit is \emph{exceptional}, despite $\lambda \approx 16.76$ not being a terribly small gauge theory 't Hooft coupling.  The accuracy becomes even better as the coupling becomes yet smaller.  

\subsection*{Strong Coupling: Evidence for the Parabolic Density}

Returning to (\ref{eqn:FilOconn}), Filev and O'Connor study the same two-matrix model  at strong coupling, after integrating over one degree of freedom to yield a one-matrix model.  They show that at large coupling, the leading order approximation to the eigenvalue density is parabolic.  This parabolic density has appeared quite often in the recent literature \cite{berenstein_multi-matrix_2009,ydri_remarks_2015,oconnor_critical_2013,delgadillo-blando_geometry_2008}.  Particularly in \cite{ydri_remarks_2015} it is remarked that the $\mathcal{N}=1^{*}$ Dijkgraaf-Vafa matrix model in the limit of large cut length supported in the real axis should reproduce the parabolic density.  Along the Hermitian slice through $\mathcal{H}$, recall that the modular solution to the theory degenerates when the two mirror cuts collide at the origin.  This seems to indicate that my construction in Chapter 4 cannot access the region capable of detecting the parabolic density.  

\vskip3ex
\noindent \textbf{However, making full use of the holomorphic nature of the $\bm{\mathcal{N}=1^{*}}$ matrix model, we may move into the bulk of $\bm{\mathcal{H}}$ and allow the two mirror cuts to elongate without colliding.  The mirror cuts can be made arbitrarily long, and then by approaching the line of degeneration, they can be made to converge to an overlapping configuration on the real axis.  Recalling that the two mirror cuts are simply translates of the actual cut, this maneuver seems to produce a branch cut of arbitrarily long length, nearly supported in the real axis with arbitrarily small imaginary part.}
\vskip3ex

In practice, the functions involved become fairly chaotic at the extreme parameter values and are quite difficult to handle computationally.  At such large 't Hooft coupling my procedure for recovering eigenvalue densities requires more involved numerical methods.  It is still a completely valid procedure, but its simple application breaks down.  I will reserve myself to approaching the real axis such that the imaginary parts of the cuts are as small as possible.  I will still be able to find convincing evidence of the parabola.  

As we've noted, eigenvalue densities in holomorphic matrix models are complex-valued, in general.  So when the cuts have non-zero imaginary part, we expect the eigenvalue densities to be complex.  As such, we must take the absolute value of the density.  Upon approaching the real axis, the imaginary part of the densities should approach zero, and so the imaginary contribution to the absolute value plays an increasingly negligible role.

\begin{figure}
\begin{center}
\includegraphics[width=12.5cm]{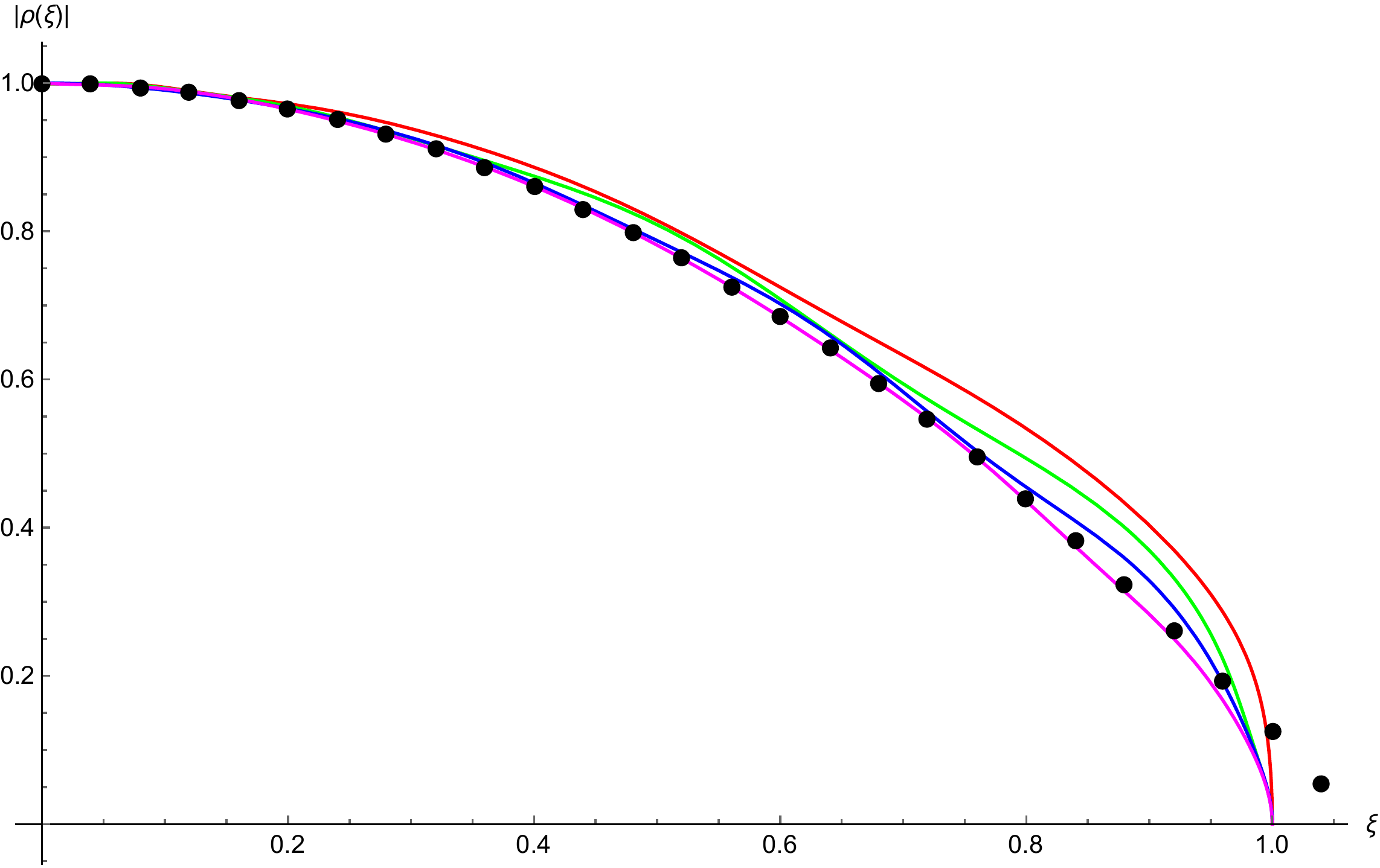}
\caption{In order of increasing coupling, the red, green, blue, and magenta plots correspond to points B, C, D, and E, respectively in Figure \ref{fig:modspplot}.  An exact parabola is shown as a black dotted line.  Notice that at the highest coupling, excellent fit to the parabola is observed over most of the support.  As described below, the fit is expected to be imperfect near the endpoints of the cut.  In order to provide a single, illustrative figure, I have scaled my eigenvalue densities to be supported on the half-length $[0,1]$ with height 1.  Undoing the scaling, my method produces normalized eigenvalue densities. \label{fig:Parab2Dens1}}
\end{center}
\end{figure}

It is clear from Figure \ref{fig:Parab2Dens1} that as the 't Hooft coupling increases, the eigenvalue density more closely resembles a parabola.  Near the cut endpoints, the fit breaks down substantially.  The reason is that an eigenvalue density must have a vertical tangent line at its endpoints.  This is certainly true of both the semi-circle, and the inverse square-root.  So the eigenvalue densities very closely resemble a parabola for most of the cut, but near the endpoints, they diverge downward to enforce the vertical tangency constraint.  Indeed, the region of disagreement near the endpoints represents a vanishingly small fraction of the total cut length at large coupling.  This disagreement of the eigenvalue density with the parabolic density was also noted in \cite{berenstein_multi-matrix_2009}.

\chapter{Conclusions}

I have reviewed and studied two main forms of geometry emerging from matrix models: an emergent algebraic curve and eigenvalue densities.  The $\mathcal{N}=1^{*}$ and $\mathcal{N}=2^{*}$ theories both give rise to an elliptic curve with generalized resolvents.  I show explicitly in the previous chapter that the two forms of emergent geometry are intimately intertwined.  Namely, the restriction of the generalized resolvent to the special cycle $\mathcal{C}^{+}$ (Figure \ref{fig:funddom1}) on the elliptic curve, allows for an exact construction of eigenvalue densities.  This unification between the two emergent structures is not particularly surprising, but is important nonetheless.

Given $\tau \in \mathbb{H}$, I attempt to reconstruct the configuration of the mirror cuts on the eigenvalue plane.  This is essentially independent of any physical theory, and was modeled on \cite{hollowood_partition_2015,dorey_exact_2002-1}.  I find there is a \emph{region of degeneration} in $\mathbb{H}$ where the reconstruction of the eigenvalue plane breaks down.  The moduli space $\mathcal{H}$ (Figure \ref{fig:modspplot}) is the compliment of the region of degeneration in $\mathbb{H}$ and the \emph{line of degeneration} is the boundary of the region of degeneration, acting as a ``line at infinity" of $\mathcal{H}$.  I find that $\text{Re}(\tau)=1/2$ is the Hermitian slice with a critical point when the two mirror branch cuts collide, consistent with \cite{hollowood_partition_2015}.  

The fundamental questions I consider in this thesis are the following.

\begin{center}
\textit{Given a Hermitian matrix model in the 't Hooft limit whose weak coupling region embeds naturally along the Hermitian slice up to the first critical point, can the geometrical solution using the elliptic curve detect the strong coupling region?  Does (the neighborhood of) the line of degeneration encode any strong coupling data?}
\end{center}

\noindent I want to emphasize that exploring the neighborhood of the line of degeneration relies on the holomorphic nature of the model; instead of terminating at the critical point, we can avoid it and move into the bulk of the moduli space.  

In the Hermitian matrix model (\ref{eqn:FilOconn}) studied in \cite{berenstein_multi-matrix_2009,ydri_remarks_2015,oconnor_critical_2013,delgadillo-blando_geometry_2008} the characteristic eigenvalue densities of a Wigner semi-circle and a parabola are discovered at weak and strong coupling, respectively.  Indeed, I show that along the Hermitian slice, $\mathcal{N}=1^{*}$ at weak coupling reproduces the Wigner semi-circular density (Figure \ref{fig:SemiCircle1}).  At strong coupling, I find encouraging evidence (Figure \ref{fig:Parab2Dens1}) that a parabolic density emerges upon approaching the line of degeneration.  This appears to be an example of a Hermitian model whose weak coupling region embeds along the Hermitian slice above the critical point, and whose strong coupling region is uncovered in a neighborhood of the line of degeneration.  If so, I have shown that an emergent elliptic curve encodes the eigenvalue densities in the model (\ref{eqn:FilOconn}) which in turn, indicate a particular non-commutative background.

I show in Chapter 5 that $\mathcal{N}=1^{*}$ restricted to the Hermitian slice through $\mathcal{H}$ encodes the same cut lengths and critical coupling $\lambda_{c}^{(1)}$ as $\mathcal{N}=2^{*}$ before the first phase transition.  Therefore, the $\mathcal{N}=2^{*}$ theory is another example of a theory which at weak coupling embeds on the Hermitian slice.  On the line of degeneration I find an infinite sequence of points $n=1,2,\ldots$ (Figure \ref{fig:modspplot}) reminiscent of the $\mathcal{N}=2^{*}$ critical points.  These points, and only these points, are the extrema of the $\mathcal{N}=1^{*}$ coupling $S(\tau)$ in the closure of $\mathcal{H}$.  In addition, approaching these points from within $\mathcal{H}$, the cut is approaching the real axis exactly with length $n|M|$ (Figure \ref{fig:CriticalPoints1} depicts the mirror cut configurations approaching these points).  These coincide with the critical cut lengths in $\mathcal{N}=2^{*}$.  The motivation is that by embedding into a holomorphic model, instead of terminating at the first critical point, one may allow the mirror cuts to move slightly off the real axis, enlarge arbitrarily, and then approach the real axis once again.  

However, only for $n=1$ does the critical 't Hooft coupling (Table \ref{Table:CritPointsTab11}) match that of $\mathcal{N}=2^{*}$.  For $n>1$, despite also being extrema of $S(\tau)$ and having the expected cut length and configuration, any possible connections to $\mathcal{N}=2^{*}$ remain a mystery.

\bibliographystyle{siam}
\bibliography{EmergentGeoHMM}



\backmatter


\end{document}